\documentclass[reqno,11pt]{amsart}

\usepackage{amssymb,enumerate}
\usepackage{charter}
\usepackage[ps2pdf,backref,colorlinks,citecolor=blue,debug]{hyperref}

\makeatletter

 \newcommand{\df}[1]{{\rmfamily\itshape\mdseries#1}}

\newcommand{\customqed}[1]{{\renewcommand{\qedsymbol}{#1}\qed}}
\newcommand{\varqed}{\customqed{\hbox{$\diamondsuit$}}}

 \theoremstyle{plain}

 \newtheorem{lemma}{Lemma}[section]
 \newtheorem{proposition}[lemma]{Proposition}
 \newtheorem{corollary}[lemma]{Corollary}

 \theoremstyle{definition}

 \newtheorem{Definition}[lemma]{Definition}

 \theoremstyle{remark}

 \newtheorem*{Notation}{Notation}
 
 \newenvironment{notation}{%
   \begin{Notation}}{\varqed\end{Notation}}

 \newtheorem{Condition}[lemma]{Condition}

 \newenvironment{condition}{%
   \begin{Condition}}{\varqed\end{Condition}}

 \newtheorem{Example}[lemma]{Example}
 \newtheorem{Examples}[lemma]{Examples}
 
 \newenvironment{example}{%
   \begin{Example}}{\varqed\end{Example}}
 \newenvironment{examples}{%
   \begin{Examples}}{\varqed\end{Examples}}

 \newtheorem{Remark}[lemma]{Remark}
 \newtheorem{Remarks}[lemma]{Remarks}
 
 \newenvironment{remark}{%
   \begin{Remark}}{\varqed\end{Remark}}

 \newcommand\ol[1]{\overline{#1}}

 \newcommand \ag{\alpha}
 \newcommand \bg{\beta}
 \newcommand \gm{\gamma}
 \newcommand \Gg{\Gamma}
 \newcommand \dg{\delta}
 \newcommand \Dg{\Delta}
 \newcommand \eps{\varepsilon}

 \newcommand \ig{\iota}
 
 \renewcommand \lg{\lambda} 
 \newcommand \Lg{\Lambda}
 \newcommand \og{\omega}
 \newcommand \Og{\Omega}
 
 \newcommand \sg{\sigma}
 \newcommand \Sg{\Sigma}

 \newcommand \tg{\tau}    

 \newcommand{\bM}{\mathbf{M}}

 \newcommand{\bU}{\mathbf{U}}

 \newcommand{\bX}{\mathbf{X}}
 \newcommand{\bY}{\mathbf{Y}}

 \newcommand{\cA}{\mathcal{A}}
 \newcommand{\cB}{\mathcal{B}}

 \newcommand{\cE}{\mathcal{E}}
 \newcommand{\cF}{\mathcal{F}}
 
 \newcommand{\cH}{\mathcal{H}}

 \newcommand{\cL}{\mathcal{L}}
 \newcommand{\cM}{\mathcal{M}}

 \newcommand{\cP}{\mathcal{P}}

 \newcommand{\cS}{\mathcal{S}}

 \newcommand{\bbB}{\mathbb{B}}
 \newcommand{\bbN}{\mathbb{N}}
 \newcommand{\bbQ}{\mathbb{Q}}
 \newcommand{\bbR}{\mathbb{R}}
 \newcommand{\bbZ}{\mathbb{Z}}

 \newcommand\setof[1]{\mathopen\{\,#1\,\mathclose\}}
 \newcommand\ang[1]{{\langle #1\rangle}}
 \newcommand\Paren[1]{{\left( #1\right)}}
 \newcommand\cei[1]{{\lceil #1\rceil}}
 \newcommand\flo[1]{{\lfloor #1\rfloor}}
 
  \renewcommand{\le}{\leqslant}
  \renewcommand{\ge}{\geqslant}
 
 \let\et=\wedge
 \let\V=\vee
 \newcommand {\sbs}{\subset}
 \newcommand {\sbsq}{\subseteq}
 
 \newcommand {\spsq}{\supseteq}
 \newcommand {\xcpt}{\mathbin{\raise0.15ex\hbox{$\smallsetminus$}}}

 \newcommand{\txt}[1]{\text{\rmfamily\mdseries\upshape{#1}}}

\newcommand{\killtext}[1]{}

\makeatother

\numberwithin{equation}{section} 

 \theoremstyle{plain}
 \newtheorem{theorem}{Theorem}
 \newtheorem{question}{Question}

\newcommand{\lea}{\stackrel{{}_+}{<}}
\newcommand{\gea}{\stackrel{{}_+}{>}}
\newcommand{\eqa}{\stackrel{{}_+}{=}}

\newcommand{\lem}{\stackrel{{}_*}{<}}
\newcommand{\gem}{\stackrel{{}_*}{>}}
\newcommand{\eqm}{\stackrel{{}_*}{=}}

\newcommand{\opint}[2]{(#1;\, #2)}
\newcommand{\clint}[2]{[#1;\, #2]}
\newcommand{\lint}[2]{[#1;\, #2)}
\newcommand{\rint}[2]{(#1;\, #2]}

\newcommand{\sqsbsq}{\sqsubseteq}
\newcommand{\ltri}{\triangleleft}

\newcommand{\dom}{\operatorname{dom}}
\newcommand{\Cl}[1]{\overline{#1}}
\newcommand{\Int}[1]{#1^{\txt{o}}}
\newcommand{\J}{J}
\renewcommand{\t}{\mathbf{t}}

\renewcommand{\d}{\mathbf{d}}
\newcommand{\m}{\mathbf{m}}
\newcommand{\En}{\operatorname{En}}
\newcommand{\Cf}{\operatorname{Cf}}

\newcommand{\Lip}{\operatorname{Lip}}
\newcommand{\id}{\operatorname{id}}
\newcommand{\Km}{\operatorname{Km}}
\newcommand{\KM} {\operatorname{KM}}

\begin{document}
  \title{Uniform test of algorithmic randomness over a general space}

   \author{Peter G\'acs}
  \address{Boston University}
  \email{gacs@bu.edu}
  \date{\today}

 \begin{abstract}
The algorithmic theory of randomness is well developed when the underlying
space is the set of finite or infinite sequences and the underlying
probability distribution is the uniform distribution or a computable
distribution.  
These restrictions seem artificial.  
Some progress has been
made to extend the theory to arbitrary Bernoulli distributions (by
Martin-L\"of), and to arbitrary distributions (by Levin).  We recall the main
ideas and problems of Levin's theory, and report further progress in the
same framework.  
The issues are the following:
{
\settowidth\leftmarginii{\hspace{0.5pc}--} 
 \begin{enumerate}[--]
  \item Allow non-compact spaces (like the space of continuous functions, 
   underlying the Brown\-ian motion).
  \item The uniform test (deficiency of randomness) $\d_{P}(x)$
(depending both on the outcome $x$ and the measure $P$) should be defined in
a general and natural way. 
  \item See which of the old results survive: 
existence of universal tests,
conservation of randomness,
expression of tests in terms of description complexity,
existence of a universal measure,
expression of mutual information as "deficiency of independence".
  \item The negative of the new randomness test is shown to be a generalization of
  complexity in continuous spaces; we show that the addition theorem
  survives.
 \end{enumerate}
  The paper's main contribution is introducing an appropriate framework for
studying these questions and related ones (like statistics for a general
family of distributions).
} 
 \end{abstract}


\maketitle

\section{Introduction}

\subsection{Problem statement}

The algorithmic theory of randomness is well developed when the underlying
space is the set of finite or infinite sequences and the underlying
probability distribution is the uniform distribution or a computable
distribution.  
These restrictions seem artificial.  
Some progress has been
made to extend the theory to arbitrary Bernoulli distributions by
Martin-L\"of in~\cite{MLof66art}, and to arbitrary distributions, by Levin
in~\cite{LevinRand73,LevinUnif76,LevinRandCons84}.
The paper~\cite{HertlingWeihrauchRand98} by Hertling and Weihrauch 
also works in general spaces, but it is restricted to computable
measures.
Similarly, Asarin's thesis~\cite{Asarin88} defines randomness for sample
paths of the Brownian motion: a fixed random process with computable
distribution.

The present paper has been inspired mainly by Levin's early 
paper~\cite{LevinUnif76} (and the much more elaborate~\cite{LevinRandCons84}
that uses different definitions): let us summarize part of the content
of~\cite{LevinUnif76}.
The notion of a constructive topological space $\bX$ and the space of
measures over $\bX$ is introduced.
Then the paper defines the notion of a uniform test.
Each test is a lower semicomputable function $(\mu,x) \mapsto f_{\mu}(x)$,
satisfying $\int f_{\mu}(x) \mu(dx) \le 1$ for each measure $\mu$.
There are also some additional conditions.
The main claims are the following.
 \begin{enumerate}[\upshape (a)]

  \item There is a universal test $\t_{\mu}(x)$, a test 
such that for each other test $f$ there is a constant $c > 0$ with
$f_{\mu}(x) \le c\cdot \t_{\mu}(x)$.
  The \df{deficiency of randomness} is defined as 
$\d_{\mu}(x) = \log\t_{\mu}(x)$. 

  \item The universal test has some strong properties of ``randomness
conservation'': these say, essentially,
that a computable mapping or a computable
randomized transition does not decrease randomness.

  \item There is a measure $M$ with the property that for
every outcome $x$ we have $\t_{M}(x) \le 1$.
In the present paper, we will call such measures \df{neutral}.

  \item\label{i.Levin.semimeasure}
Semimeasures (semi-additive measures)
are introduced and it is shown that there is a lower semicomputable
semimeasure that is neutral (so we can assume that the $M$ introduced above
is lower semicomputable).

  \item Mutual information $I(x : y)$
is defined with the help of (an appropriate
version of) Kolmogorov complexity, between outcomes $x$ and $y$.
It is shown that $I(x : y)$ is essentially equal to $\d_{M \times M}(x,y)$.
This interprets mutual information as a kind of ``deficiency of independence''.

 \end{enumerate}
This impressive theory leaves a number of issues unresolved:
  \begin{enumerate}[\upshape 1.]

   \item The space of outcomes is restricted to be a compact topological
space, moreover, a particular compact space: the set of 
sequences over a finite alphabet (or, implicitly in~\cite{LevinRandCons84},
a compactified infinite alphabet).
However, a good deal of modern probability theory happens over
spaces that are not even locally compact:
for example, in case of the Brownian motion, over the
space of continuous functions.

   \item The definition of a uniform randomness test includes some conditions
(different ones in~\cite{LevinUnif76} and 
in~\cite{LevinRandCons84}) that seem somewhat arbitrary.

   \item No simple expression is known for the general
universal test in terms of description complexity.
Such expressions are nice to have if they are available.

  \end{enumerate}

\subsection{Content of the paper}

The present paper intends to carry out as much of Levin's program as seems
possible after removing the restrictions.
It leaves a number of questions open, but we feel that they are worth
to be at least formulated.
A fairly large part of the paper 
is devoted to the necessary conceptual machinery.
Eventually, this
will also allow to carry further some other initiatives started in the
works~\cite{MLof66art} and~\cite{LevinRand73}: the study of tests that test
nonrandomness with respect to a whole class of measures (like the Bernoulli
measures).

Constructive analysis has been developed by several authors,
converging approximately on the same concepts.
We will make use of a simplified  version of the theory introduced  
in~\cite{WeihrauchComputAnal00}.
As we have not found a constructive measure theory in the
literature fitting our purposes, we will develop this theory here,
over (constructive) complete separable metric spaces.
This generality is well supported by standard results in measure
theoretical probability, and is
sufficient for constructivizing a large part of current probability
theory.

The appendix recalls some of the needed topology, measure theory
and constructive analysis.
Constructive measure theory is introduced in Section~\ref{s.constr-meas}.

Section~\ref{s.unif-test} introduces uniform randomness tests.
It proves the existence of universal uniform tests, under a reasonable
assumption about the topology (``recognizable Boolean inclusions'').
Then it proves conservation of randomness.

Section~\ref{s.complexity} explores the relation between description 
(Kolmogorov) complexity and uniform randomness tests.
After extending randomness tests 
over non-normalized measures, its negative logarithm will be seen
as a generalized description complexity.

The rest of the section explores the extent to which the old results
characterizing a random infinite string by the description complexity of
its segments can be extended to the new setting.
We will see that the simple formula working for computable measures over
infinite sequences does not generalize.
However, still rather simple formulas are available in some cases: namely,
the discrete case with general measures, and a space allowing a certain
natural cell decomposition, in case of computable measures.

Section~\ref{s.neutral} proves Levin's theorem about the existence of a
neutral measure, for compact spaces.
Then it shows that the result does not generalize to non-compact spaces,
not even to the discrete space.
It also shows that with our definition of tests, the neutral measure cannot
be chosen semicomputable, even in the case of the discrete space
with one-point compactification.

Section~\ref{s.rel-entr} takes up the idea of viewing 
the negative logarithm of a randomness test as generalized description
complexity.
Calling this notion \df{algorithmic entropy}, this section explores its
information-theoretical properties.
The main result is a (nontrivial)
generalization of the addition theorem of prefix
complexity (and, of course, classical entropy) to the new setting.

\subsection{Some history}

Attempts to define randomness rigorously have
a long but rather sparse history starting with von Mises and continuing
with Wald, Church, Ville.
Kolmogorov's work in this area inspired Martin-L\"of whose
paper~\cite{MLof66art} introduces the notion of randomness used here.

Description complexity has been introduced independently by
Solomonoff, Kolmogorov and Chaitin.
Prefix complexity has been introduced independently by Levin and Chaitin.
See~\cite{LiViBook97} for a discussion of priorities and contributions.
The addition theorem (whose generalization is given here) has been proved
first for Kolmogorov complexity, with a logarithmic error term, by
Kolmogorov and Levin.
For the prefix complexity its present form has been proved
jointly by Levin and G\'acs in~\cite{GacsSymm74}, and
independently by Chaitin in~\cite{Chaitin75}.

In his PhD thesis, Martin-L\"of also characterized randomness of finite 
sequences via their complexity.
For infinite sequences, complete characterizations of their
randomness via the complexity of their segments were given
by Levin in~\cite{LevinRand73}, by Schnorr
in~\cite{Schnorr73} and in~\cite{Chaitin75} (attributed).
Of these, only Levin's result is formulated for general computable measures:
the others apply only to coin-tossing.
Each of these works uses a different variant of description complexity.
Levin uses monotone complexity and the logarithm of the universal
semicomputable measure (see~\cite{GacsRel83} for the difficult proof that 
these two complexities are different).
Schnorr uses ``process complexity'' (similar to monotone 
complexity) and prefix complexity.
The work~\cite{GacsExact80} by the present author gives
characterizations using the original Kolmogorov
complexity (for general computable measures).

Uniform tests over the space of infinite sequences, 
randomness conservation and neutral measures
were introduced in Levin's work~\cite{LevinUnif76}.
The present author could not verify every result in that paper (which
contains no proofs); he reproduced most of them with a changed definition
in~\cite{GacsExact80}.
A universal uniform test with yet another definiton appeared
in~\cite{LevinRandCons84}.
In this latter work, ``information conservation'' is a central tool 
used to derive several results in logic.
In the constellation of Levin's concepts, information conservation
becomes a special case of randomness conservation.
We have not been able to reproduce this exact relation with our definition
here.

The work~\cite{GacsBoltzmann94} is based on the observation that
Zurek's idea on ``physical'' entropy and the ``cell volume'' approach of
physicists to the definition of entropy can be unified: Zurek's entropy
can be seen as an approximation of the limit arising in a
characterization of a randomness test by complexity.
The author discovered in this same paper that the negative logarithm of a
general randomness test can be seen as a generalization of complexity.
He felt encouraged by the discovery of the generalized addition theorem
presented here.

The appearence of other papers in the meantime
(including~\cite{HertlingWeihrauchRand98}) convinced the author that
there is no accessible and detailed reference work on
algorithmic randomness for general measures and general spaces, 
and a paper like the present one, developing the foundations, is needed.
(Asarin's thesis~\cite{Asarin88} does develop the theory of randomness for
the Brownian motion.
It is a step in our direction in the sense that the space is not compact,
but it is all done for a single explicitly given computable measure.)

We do not advocate the uniform randomness test proposed here
as necessarily the ``definitive'' test concept.
Perhaps a good argument can be found for some additional conditions,
similar to the ones introduced by Levin, providing additional structure
(like a semicomputable neutral measure) while preserving naturalness
and the attractive properties presented here.

\subsection{Notation for the paper}

(Nothing to do with the formal concept of ``notation'', introduced
later in the section on constructive analysis.)
The sets of natural numbers, integers, rational numbers, real numbers
and complex numbers
will be denoted respectively by $\bbN, \bbZ, \bbQ, \bbR$.
The set of nonnegative real numbers will be denoted by $\bbR_{+}$.
The set of real numbers with $-\infty,\infty$ added (with the appropriate
topology making it compact) will be denoted by $\ol\bbR$.
We use $\et$ and $\V$ to denote $\min$ and $\max$, further
 \[
   |x|^{+} = x\V 0,\quad |x|^{-} = |-x|^{+}
 \]
for real numbers $x$.
We partially follow~\cite{WeihrauchComputAnal00},
\cite{BrattkaComputTopStruct03} and~\cite{HertlingWeihrauchRand98}.
In particular, adopting the notation of~\cite{WeihrauchComputAnal00},
we denote intervals of the real line as follows, (to avoid the conflict
with the notation of a pair $(a,b)$ of objects).
 \[
   \clint{a}{b} = \setof{x : a \le x \le b}, \quad 
   \opint{a}{b}  = \setof{x : a < x < b},
\quad \lint{a}{b} = \setof{x : a \le x < b}.
 \]
If $X$ is a set then $X^{*}$ is
the set of all finite strings made up of elements of $X$, including the 
``empty string'' $\Lg$.
We denote by $X^{\og}$ the set of all infinite sequences of elements of
$X$.
If $A$ is a set then $1_{A}(x)$ is its indicator function, defined to
be 1 if $x\in A$ and to 0 otherwise.
For a string $x$, its length is $|x|$, and
 \[
  x^{\le n} = (x(1),\dots,x(n)).
 \]
The relations
 \[
  f \lea g,\quad f\lem g
 \]
mean inequality to within an additive constant and multiplicative constant
respectively.
The first is equivalent to $f \le g + O(1)$, the second to $f = O(g)$. 
The relation $f\eqm g$ means $f \lem g$ and $f \gem g$.

 Borrowing from~\cite{PollardUsers01}, for
a function $f$ and a measure $\mu$, we will use the notation
 \[
      \mu f = \int f(x)\mu(dx),
\quad \mu^{y} f(x, y) = \int f(x,y)\mu(dy).
 \]

\section{Constructive measure theory}\label{s.constr-meas}

The basic concepts and results of measure theory are recalled in
Section~\ref{s.measures}.
For the theory of measures over metric spaces, see
Subsection~\ref{ss.measure-metric}.
We introduce a certain fixed, enumerated sequence of Lipschitz functions
that will be used frequently.
Let $\cF_{0}$ be the set of functions
of the form $g_{u,r,1/n}$ where $u \in D$, $r\in \bbQ$, $n = 1, 2, \dots$,  
and
 \begin{equation*}
  g_{u,r,\eps}(x) = |1 - |d(x, u) - r|^{+}/\eps|^{+}
 \end{equation*}
 is a continuous function that is $1$ in the ball
$B(u,r)$, it is 0 outside $B(u, r+\eps)$, and takes intermediate
values in between.
Let 
 \begin{equation}\label{e.bd-Lip-seq}
  \cE = \{g_{1}, g_{2}, \dots \}
 \end{equation}
 be the smallest set of functions containing $\cF_{0}$
and the constant 1, and closed under $\V$, $\et$ and rational linear
combinations.  
The following construction will prove useful later.

 \begin{proposition}\label{p.bd-Lip-set}
All bounded continuous functions can be 
obtained as the limit of an increasing sequence of functions from
the enumerated countable set $\cE$ of bounded computable
Lip\-schitz functions introduced in~\eqref{e.bd-Lip-seq}.
 \end{proposition}
The proof is routine.

\subsection{Space of measures}

Let $\bX = (X, d, D, \ag)$ be a computable metric space.
In Subsection~\ref{ss.measure-metric}, the space
$\cM(\bX)$ of measures over $\bX$ is defined, along with a natural 
enumeration $\nu = \nu_{\cM}$ for a subbase $\sg = \sg_{\cM}$
of the weak topology.
This is a constructive topological space $\bM$ which can be metrized by
introducing, as in~\ref{sss.Prokh}, the 
\df{Prokhorov distance} $p(\mu, \nu)$: 
the infimum of all those $\eps$ for which, for all Borel sets $A$ we have
$\mu(A) \le \nu(A^{\eps}) + \eps$, 
where $A^{\eps} = \setof{x : \exists y\in A\; d(x, y) < \eps}$.
Let $D_{\bM}$ be the set of
those probability measures that are concentrated on finitely many points of
$D$ and assign rational values to them.
Let $\ag_{\bM}$ be a natural enumeration of $D_{\bM}$.
Then 
 \begin{equation}\label{e.metric-measures}
   (\cM, p, D_{\bM}, \ag_{\bM})
 \end{equation}
is a computable metric space whose constructive topology is equivalent to
$\bM$.
Let $U=B(x, r)$ be one of the balls in $\bX$, 
where $x\in D_{\bX}$, $r \in \bbQ$.
The function $\mu \mapsto \mu(U)$ is typically not computable, 
not even continuous.
For example, if $\bX=\bbR$ and $U$ is the open interval $\opint{0}{1}$, 
the sequence of probability measures $\dg_{1/n}$ (concentrated on $1/n$)
converges to $\dg_{0}$, but $\dg_{1/n}(U)=1$, and $\dg_{0}(U)=0$.
The following theorem shows that the situation is better with
$\mu \mapsto \mu f$ for computable $f$:

 \begin{proposition}\label{p.computable-integral}
Let $\bX = (X, d, D, \ag)$ be a computable metric space, and let
$\bM = (\cM(\bX), \sg, \nu)$ be the effective topological space of
probability measures over $\bX$.
If function $f : \bX \to \bbR$ is bounded and computable
then $\mu \mapsto \mu f$ is computable.
 \end{proposition}
 \begin{proof}[Proof sketch]
To prove the theorem for bounded Lip\-schitz functions, we can invoke
the Strassen coupling theorem~\ref{p.coupling}.

The function $f$ can be obtained as a limit of a computable
monotone increasing sequence of computable Lip\-schitz functions $f^{>}_{n}$,
and also as a limit of a computable monotone decreasing 
sequence of computable Lip\-schitz functions $f^{<}_{n}$.
In step $n$ of our computation of $\mu f$,
we can approximate $\mu f^{>}_{n}$ from above 
to within $1/n$, and $\mu f^{<}_{n}$ from below to within $1/n$.
Let these bounds be $a^{>}_{n}$ and $a^{<}_{n}$.
To approximate $\mu f$ to within $\eps$, 
find a stage $n$ with $a^{>}_{n} - a^{<}_{n} +2/n < \eps$.
 \end{proof}

\subsection{Computable measures and random 
transitions}\label{ss.computable-trans}
A measure $\mu$ is called \df{computable} if it is a computable element of the
space of measures.
Let $\{g_{i}\}$ be the set of bounded Lip\-schitz functions over $X$
introduced  in~\eqref{e.bd-Lip-seq}.

 \begin{proposition}\label{p.computable-meas-crit}
Measure $\mu$ is computable if and only if so is the function
$i \mapsto \mu g_{i}$.
 \end{proposition}
 \begin{proof}
The ``only if'' part follows from Proposition~\ref{p.computable-integral}.
For the ``if'' part, note that in order to 
trap $\mu$ within some Prokhorov neighborhood of size $\eps$,
it is sufficient to compute $\mu g_{i}$ within a small
enough $\dg$, for all $i\le n$ for a large enough $n$.
 \end{proof}

 \begin{example}
Let our probability space be the set $\bbR$ of real numbers with its
standard topology.
Let $a < b$ be two computable real numbers.
Let $\mu$ be the probability
distribution with density function
 $f(x) = \frac{1}{b-a}1_{\clint{a}{b}}(x)$ 
(the uniform distribution over the interval $\clint{a}{b}$).
Function $f(x)$ is not computable, since it is not even continuous.
However, the measure $\mu$ is computable: indeed,
$\mu g_{i} = \frac{1}{b-a} \int_{a}^{b} g_{i}(x) dx$ is a computable
sequence, hence 
Proposition~\ref{p.computable-meas-crit} implies that $\mu$ is computable.
 \end{example}

The following theorem compensates somewhat for the fact
mentioned earlier, that the
function $\mu \mapsto \mu(U)$ is generally not computable.

 \begin{proposition}
Let $\mu$ be a finite computable measure.
Then there is a computable map $h$ with the property that for every
bounded computable function $f$ with $|f| \le 1$ with
the property $\mu(f^{-1}(0))=0$,
if $w$ is the name of $f$ then $h(w)$ is the
name of a program computing the value $\mu\setof{x: f(x) < 0}$.
 \end{proposition}
 \begin{proof}
Straightforward.
 \end{proof}

 \begin{remark}
Suppose that there is a computable function that for each $i$ computes a
Cauchy sequence $j \mapsto m_{i}(j)$ with the property that for 
$i < j_{1} < j_{2}$ we have $|m_{i}(j_{1})-m_{i}(j_{2})| < 2^{-j_{1}}$, and
that for all $n$, there is a measure $\nu$ with the property that
for all $i \le n$, $\nu g_{i} = m_{i}(n)$.
Is there a measure $\mu$ with the property that for each $i$ we have
$\lim_{j} m_{i}(j) = \mu g_{i}$?
Not necessarily, if the space is not compact.
For example, let $X = \{1,2,3,\dots\}$ with the discrete topology.
The sequences $m_{i}(j) = 0$ for $j > i$ satisfy these conditions, but
they converge to the measure 0, not to a probability measure.
To guarantee that the sequences $m_{i}(j)$ indeed define a probability
measure, progress must be made, for example, in terms of the narrowing of
Prokhorov neighborhoods.
 \end{remark}

Let now $\bX,\bY$ be computable metric spaces.
They give rise to measurable spaces with $\sg$-algebras $\cA, \cB$
respectively.
Let $\Lg = \setof{\lg_{x} : x \in X}$ be a probability kernel from $X$ to
$Y$ (as defined in Subsection~\ref{ss.transitions}).
Let $\{g_{i}\}$ be the set of bounded Lip\-schitz functions over $Y$ 
introduced in~\eqref{e.bd-Lip-seq}.
To each  $g_{i}$, the kernel assigns a (bounded) measurable function 
 \[
   f_{i}(x) = (\Lg g_{i})(x) =  \lg_{x}^{y} g_{i}(y).
 \]
We will call $\Lg$ \df{computable} if so is the assignment 
$(i, x) \mapsto f_{i}(x)$.
In this case, of course, each function $f_{i}(x)$ is continuous.
The measure $\Lg^{*}\mu$ is determined by the values 
$\Lg^{*} g_{i} = \mu (\Lg g_{i})$, which are computable from $(i, \mu)$ and
so the function $\mu \mapsto \Lg^{*}\mu$ is computable.

 \begin{example}\label{x.computable-determ-trans}
A computable function $h : X \to Y$ defines an operator 
$\Lg_{h}$ with $\Lg_{h} g = g \circ h$ (as in Example~\ref{x.determ-trans}).
This is a deterministic computable transition, in which
$f_{i}(x) = (\Lg_{h} g_{i})(x) = g_{i}(h(x))$ is, 
of course, computable from $(i,x)$.
We define $h^{*}\mu = \Lg_{h}^{*}\mu$.
 \end{example}

\subsection{Cells}\label{ss.cells}

As pointed out earlier, it is not convenient to define
a measure $\mu$ constructively starting from $\mu(\Gg)$ for open cells
$\Gg$. 
The reason is that no matter how we fix $\Gg$, the function
$\mu \mapsto \mu(\Gg)$ is typically not computable.
It is better to work with bounded computable functions, since for such
a function $f$, the correspondence $\mu \mapsto \mu f$ is computable.

Under some special conditions, we will still get ``sharp'' cells.
Let $f$ be a bounded computable function over $\bX$, let
$\ag_{1}<\dots<\ag_{k}$ be rational numbers,
and let $\mu$ be a computable measure with the property
that $\mu f^{-1}(\ag_{j})=0$ for all $j$.
In this case, we will say that $\ag_{j}$ are \df{regular points} of $f$
with respect to $\mu$.
Let $\ag_{0}=-\infty$, $\ag_{k+1}=\infty$, and for $j=0,\dots,k$, let
Let $U_{j} = f^{-1}((j,j+1))$.
The sequence of disjoint 
r.e.~open sets $(U_{0},\dots,U_{k})$ will be called the
\df{partition generated by} $f,\ag_{1},\dots,\ag_{k}$.
(Note that this sequence is not a partition in the sense of
$\bigcup_{j}U_{j}=\bX$, since the boundaries of the sets are left out.)
If we have several partitions $(U_{i0},\dots,U_{i,k})$, 
generated by different functions $f_{i}$ ($i=1,\dots,m$)
and different regular cutoff sequences $(\ag_{ij}: j=1,\dots,k_{i})$, 
then we can form a new partition generated by all possible intersections
 \[
  V_{j_{1},\dots,j_{n}} = U_{1,j_{1}}\cap \dots \cap U_{m,j_{m}}.
 \]
A partition of this kind will be called a \df{regular partition}.
The sets $V_{j_{1},\dots,j_{n}}$ will be called the \df{cells} of this
partition. 

 \begin{proposition}\label{p.reg-partit-meas-cptable}
In a regular partition as given above, 
the values $\mu V_{j_{1},\dots,j_{n}}$ are computable
from the names of the functions $f_{i}$ and the cutoff points $\ag_{ij}$.
 \end{proposition}
 \begin{proof}
Straightforward.
 \end{proof}

Assume that a computable sequence of functions 
$b_{1}(x),b_{2}(x),\dots$ over $X$ is given,
with the property that for every pair 
$x_{1},x_{2}\in X$ with $x_{1}\ne x_{2}$, there is a $j$
with $b_{j}(x_{1})\cdot b_{j}(x_{2}) < 0$.
Such a sequence will be called a \df{separating sequence}.
Let us give the correspondence between the set $\bbB^{\og}$
of infinite binary sequences and elements of the set
 \[
  X^{0} = \setof{x\in X: b_{j}(x)\ne 0,\;j=1,2,\dots}.
 \]
For a binary string 
$s_{1}\dotsm s_{n} = s\in\bbB^{*}$, let
 \[
   \Gg_{s}
 \]
be the set of elements of $X$ with the property that for
$j=1,\dots,n$, if $s_{j}=0$ then $b_{j}(\og) < 0$, otherwise
$b_{j}(\og)>0$.
This correspondence has the following properties.
 \begin{enumerate}[(a)]
  \item $\Gg_{\Lg}=X$.
  \item For each $s\in \bbB$, the sets $\Gg_{s0}$ and 
$\Gg_{s1}$ are disjoint and their union is contained in $\Gg_{s}$.
  \item For $x\in X^0$, we have $\{x\} = \bigcap_{x\in\Gg_{s}} \Gg_{s}$.
 \end{enumerate}
If $s$ has length $n$ then $\Gg_{s}$ will be called a \df{canonical
$n$-cell}, or simply canonical cell, or $n$-cell.
From now on, whenever $\Gg$ denotes a subset of $X$, it means a
canonical cell.
We will also use the notation
 \[
   l(\Gg_{s})=l(s).
 \]
The three properties above say that if we restrict ourselves to the set
$X^0$ then the canonical cells behave somewhat like binary subintervals:
they divide $X^0$ in half, then each half again in half, etc.  
Moreover, around each point, these canonical cells become ``arbitrarily
small'', in some sense (though, they may not be a basis of neighborhoods).
It is easy to see that if $\Gg_{s_1},\Gg_{s_2}$ are two canonical
cells then they either are disjoint or one of them contains the other.
If $\Gg_{s_1}\sbs\Gg_{s_2}$ then $s_2$ is a prefix of $s_1$.
If, for a moment, we write $\Gg^0_s=\Gg_s\cap X^0$ then we have the
disjoint union $\Gg^0_s=\Gg^0_{s0}\cup\Gg^0_{s1}$.
For an $n$-element binary string $s$, for $x \in \Gg_{s}$, we will write 
 \[
    \mu(s) = \mu(\Gg_{s}).
 \]
  Thus, for elements of $X^0$, we can talk about the $n$-th bit $x_n$
of the description of $x$: it is uniquely determined.
The $2^n$ cells (some of them possibly empty)
of the form $\Gg_s$ for $l(s)=n$ form a partition
 \[
  \cP_n
 \]
  of $X^0$.

 \begin{examples}\label{x.cells}\
 \begin{enumerate}[\upshape 1.]
  \item If $\bX$ is the set of infinite binary sequences with its usual
topology, the functions $b_{n}(x) = x_{n}-1/2$ generate the usual
cells, and $\bX^{0}=\bX$.
  \item If $\bX$ is the interval $\clint{0}{1}$, let 
$b_{n}(x) = -\sin(2^{n}\pi x)$.
Then cells are open 
intervals of the form $\opint{k\cdot 2^{-n}}{(k+1)\cdot 2^{n}}$,
the correspondence between infinite binary strings
and elements of $X^0$ is just the usual representation of $x$ as the
binary decimal string $0.x_{1}x_{2}\dots$.
 \end{enumerate}
 \end{examples}

When we fix canonical cells,  
we will generally assume that the partition chosen is also
``natural''.
The bits $x_1,x_2,\ldots$ could contain information about the
point $x$ in decreasing order of importance from a macroscopic
point of view. 
For example, for a container of gas, the first few bits may
describe, to a reasonable degree of precision, the amount of gas in
the left half of the container, the next few bits may describe the
amounts in each quarter, the next few bits may describe the
temperature in each half, the next few bits may describe again the
amount of gas in each half, but now to more precision, etc.
From now on, whenever $\Gg$ denotes a subset of $X$, it means a
canonical cell.
From now on, for elements of $X^0$, we can talk about the $n$-th
bit $x_n$ of the description of $x$: it is uniquely determined.

The following observation will prove useful.

 \begin{proposition}\label{p.compact-cell-basis}
Suppose that the space $\bX$ is compact and we have a separating
sequence $b_{i}(x)$ as given above.
Then the cells $\Gg_{s}$ form a basis of the space $\bX$.
 \end{proposition}
 \begin{proof}
We need to prove that for every ball $B(x,r)$, there is a cell
$x\in\Gg_{s}\sbs B(x,r)$.
Let $C$ be the complement of $B(x,r)$.
For each point $y$ of $C$, there is an $i$ such that
$b_{i}(x)\cdot b_{i}(y) < 0$.
In this case, let $J^{0} = \setof{z: b_{i}(z) < 0}$,
$J^{1} = \setof{z: b_{i}(z) > 0}$.
Let $J(y)=J^{p}$ such that $y\in J^{p}$.
Then $C \sbs \bigcup_{y} J(y)$, and compactness implies that there is a
finite sequence $y_{1},\dots,y_{k}$ with 
$C \sbs \bigcup_{j=1}^{k} J(y_{j})$.
Clearly, there is a cell 
$x \in \Gg_{s} \sbs B(x,r) \xcpt \bigcup_{j=1}^{k} J(y_{j})$.
 \end{proof}

 \section{Uniform tests}\label{s.unif-test}

\subsection{Universal uniform test}

Let $\bX = (X, d, D, \ag)$ be a computable metric space, and let
$\bM = (\cM(\bX), \sg, \nu)$ be the constructive topological space of
probability measures over $\bX$.
A \df{randomness test} is a function $f : \bM \times \bX \to \ol\bbR$
with the following two properties.

 \begin{condition}\label{cnd.test}\
  \begin{enumerate}[\upshape 1.]
   \item The function $(\mu, x) \mapsto f_{\mu}(x)$ is lower
semicomputable.
(Then for each $\mu$, the integral $\mu f_{\mu} = \mu^{x} f_{\mu}(x)$
exists.)
   \item\label{i.test.integr} $\mu f_{\mu} \le 1$.
  \end{enumerate}
 \end{condition}

The value $f_{\mu}(x)$ is intended to quantify
the nonrandomness of the outcome $x$ with respect to the probability
measure $\mu$.
The larger the values the less random is $x$.
Condition~\ref{cnd.test}.\ref{i.test.integr} guarantees that the
probability of those outcomes whose randomness is $\ge m$ is at most
$1/m$.
The definition of tests is in the spirit of Martin-L\"of's tests.
The important difference is in the semicomputability condition:
instead of restricting the measure $\mu$ to be computable, we require the
test to be lower semicomputable also in its argument $\mu$.

Just as with Martin-L\"of's tests, we want to find a universal test;
however, we seem to need a condition on the space $\bX$.
Let us say that a sequence $i \mapsto U_{i}$ of sets has \df{recognizable
Boolean inclusions} if the set
 \[
   \setof{ (S, T) : S, T \txt{ are finite, } 
  \bigcap_{i \in S} U_{i} \sbs \bigcup_{j \in T} U_{j}}
 \]
is recursively enumerable.
We will say that a computable metric space has recognizable Boolean inclusions
if this is true of the enumerated basis consisting of balls of the form
$B(x, r)$ where $x \in D$ and $r > 0$ is a rational number.

It is our conviction that the important metric spaces studied in
probability theory have recognizable Boolean inclusions, 
and that proving this in each
individual case should not be too difficult.
For example, it does not seem difficult to prove this for the space
$C\clint{0}{1}$ of Example~\ref{x.cptable-metric-{0}{1}}, with the set of
rational piecewise-linear functions chosen as $D$.
But, we have not carried out any of this work!

 \begin{theorem}\label{t.univ-unif}
Suppose that the metric space $\bX$ has recognizable
Boolean inclusions.
Then there is a universal test, that is a test $\t_{\mu}(x)$ with the
property 
that for every other test $f_{\mu}(x)$ there is a constant $c_{f} > 0$ with
$c_{f} f_{\mu}(x) \le \t_{\mu}(x)$.
 \end{theorem}
 \begin{proof}\
\begin{enumerate}[1.]

  \item\label{univ-unif.g'}
We will show that there is a mapping that to each name $u$ of a lower
semicomputable function $(\mu, x) \mapsto g(\mu, x)$ assigns the name of a
lower semicomputable function 
$g'(\mu, x)$ such that $\mu^{x} g'(\mu,x) \le 1$, and
if $g$ is a test then $g'=g$.

To prove the statement, let us represent the space $\bM$ rather as 
 \begin{equation}\label{e.metric-measures-again}
  \bM = (\cM(\bX), p, D, \ag_{\bM}),
 \end{equation}
as in~\eqref{e.metric-measures}.
Since $g(\mu, x)$ is lower semicomputable, there is a computable sequence of
basis elements $U_{i} \sbs \bM$ and $V_{i} \sbs \bX$ 
and rational bounds $r_{i}$ such that
 \[
   g(\mu, x) = \sup_{i}\; r_{i} 1_{U_{i}}(\mu)1_{V_{i}}(x).
 \]
Let $h_{n}(\mu, x) = \max_{i \le n} r_{i} 1_{U_{i}}(\mu)1_{V_{i}}(x)$.
Let us also set $h_{0}(\mu, x) = 0$.
Our goal is to show that
the condition $\forall \mu\; \mu^{x} h_{n}(\mu, x) \le 1$ is decidable.
If this is the case then we will be done.
Indeed, we can define $h'_{n}(\mu,x)$ recursively as follows.
Let $h'_{0}(\mu, x)=0$.
Assume that $h'_{n}(\mu,x)$ has been defined already.
If $\forall\mu\; \mu^{x} h_{n+1}(\mu,x) \le 1$ then
$h'_{n+1}(\mu,x) = h_{n+1}(\mu,x)$; otherwise, it is $h'_{n}(\mu,x)$.
The function $g'(\mu, x) = \sup_{n} h'_{n}(\mu,x)$ clearly satisfies our
requirements.

We proceed to prove the decidability of the condition 
 \begin{equation}\label{e.finite-test-cond}
  \forall \mu\;  \mu^{x} h_{n}(\mu, x) \le 1.
 \end{equation}
The basis elements $V_{i}$ can be taken as
balls $B(q_{i},\dg_{i})$ for a computable sequence $q_{i} \in D$ and
computable sequence of rational numbers $\dg_{i}>0$.
Similarly, the basis element $U_{i}$ is the set of measures that is a ball
$B(\sg_{i}, \eps_{i})$, in the metric
space~\eqref{e.metric-measures-again}.
Here, using notation~\eqref{e.finite-Prokhorov},
$\sg_{i}$ is a measure concentrated on a finite set $S_{i}$.
According to Proposition~\ref{p.simple-Prokhorov-ball},
the ball $U_{i}$ is the set of measures $\mu$ satisfying the inequalities
 \[
   \mu(A^{\eps_{i}}) > \sg_{i}(A) - \eps_{i}
 \]
for all $A \sbs S_{i}$.
For each $n$, consider the finite set of balls 
 \[
  \cB_{n} = \setof{B(q_{i},\dg_{i}) : i \le n} \cup 
   \setof{B(s, \eps_{i}) : i \le n,\; s \in S_{i}}.
 \]
Consider all sets of the form 
 \[
     U_{A,B} = \bigcap_{U \in A} U \xcpt \bigcup_{U \in B} U
 \]
for all pairs of sets $A, B \sbs \cB_{n}$.
These sets are all finite intersections of balls or complements of
balls from the finite set $\cB_{n}$ of balls.
The space $\bX$ has
recognizable Boolean inclusions, so it is decidable which of these sets
$U_{A,B}$ are nonempty.
The condition~\eqref{e.finite-test-cond} can be formulated as a
Boolean formula involving linear inequalities with rational coefficients,
for the variables $\mu_{A,B}=\mu(U_{A,B})$, for those $A,B$ with
$U_{A,B}\ne\emptyset$.
The solvability of such a Boolean condition can always be decided.

 \item\label{univ-unif.end}
  Let us enumerate all lower semicomputable functions $g_{u}(\mu, x)$ for
all the names $u$.
Without loss of generality, assume these names to be natural numbers, 
and form the functions $g'_{u}(\mu,x)$ according to the
assertion~\ref{univ-unif.g'} above.
The function $t = \sum_{u} 2^{-u-1} g'_{u}$ will be the desired universal
test. 
\end{enumerate}
 \end{proof}

From now on, when referring to randomness tests, we will always assume that
our space $\bX$ has recognizable Boolean inclusions and hence has a
universal test.
We fix a universal test $\t_{\mu}(x)$, and call the function
 \[
   \d_{\mu}(x) = \log \t_{\mu}(x).
 \]
the \df{deficiency of randomness} of $x$ with respect to $\mu$.
We call an element $x\in X$ \df{random} with respect to $\mu$
if $\d_{\mu}(x) < \infty$.

 \begin{remark}\label{r.cond-test}
Tests can be generalized to include an arbitrary parameter $y$:
we can talk about the universal test 
 \[
  \t_{\mu}(x \mid y),
 \]
where $y$ comes from some constructive topological space $\bY$.
This is a maximal (within a multiplicative constant) lower semicomputable
function $(x,y,\mu) \mapsto f(x,y,\mu)$ with the property
$\mu^{x} f(x,y,\mu) \le 1$.
 \end{remark}

\subsection{Conservation of randomness}

For $i=1,0$, let
$\bX_{i} = (X_{i}, d_{i}, D_{i}, \ag_{i})$ be computable metric spaces, 
and let $\bM_{i} = (\cM(\bX_{i}), \sg_{i}, \nu_{i})$ 
be the effective topological space of
probability measures over $\bX_{i}$.
Let $\Lg$ be a computable probability kernel from $\bX_{1}$ to $\bX_{0}$ as
defined in Subsection~\ref{ss.computable-trans}.
In the following theorem, the same notation $\d_{\mu}(x)$ will
refer to the deficiency of randomness with respect to two different spaces,
$\bX_{1}$ and $\bX_{0}$, but this should not cause confusion.
Let us first spell out the conservation theorem before interpreting it.

 \begin{theorem}\label{p.conservation}
For a computable probability kernel $\Lg$ from $\bX_{1}$ to $\bX_{0}$,
we have 
 \begin{equation}\label{e.conservation}
  \lg_{x}^{y} \t_{\Lg^{*}\mu}(y) \lem \t_{\mu}(x).
 \end{equation}
 \end{theorem}
 \begin{proof}
Let $\t_{\nu}(x)$ be the universal test over $\bX_{0}$.
The left-hand side of~\eqref{e.conservation} can be written as
 \[
  u_{\mu} = \Lg \t_{\Lg^{*}\mu}.
 \]
According to~\eqref{e.Lg-Lg*}, we have 
$\mu u_{\mu} = (\Lg^{*}\mu) \t_{\Lg^{*}\mu}$ which is $\le 1$ since $\t$ is
a test.
If we show that $(\mu,x) \mapsto u_{\mu}(x)$ is lower semicomputable
then the universality of $\t_{\mu}$ will imply $u_{\mu} \lem \t_{\mu}$.

According to Proposition~\ref{p.lower-semi-as-limit}, as
a lower semicomputable function, $\t_{\nu}(y)$ can be written as 
$\sup_{n} g_{n}(\nu, y)$, where $(g_{n}(\nu, y))$ is a computable sequence
of computable functions.
We pointed out in Subsection~\ref{ss.computable-trans} that the function 
$\mu \mapsto \Lg^{*}\mu$ is computable.
Therefore the function $(n, \mu, x) \mapsto g_{n}(\Lg^{*}\mu, f(x))$ 
is also a computable.
So, $u_{\mu}(x)$ is the supremum of a computable sequence of computable
functions and as such, lower semicomputable.
 \end{proof}

It is easier to interpret the theorem first in the special case when
$\Lg = \Lg_{h}$ for a computable function $h : X_{1} \to X_{0}$,
as in Example~\ref{x.computable-determ-trans}.
Then the theorem simplifies to the following.

 \begin{corollary}\label{c.conservation-determ}
For a computable function $h : X_{1} \to X_{0}$, we have
$\d_{h^{*}\mu}(h(x)) \lea \d_{\mu}(x)$.
 \end{corollary}

Informally, this says that if $x$ is random with respect to $\mu$ in
$\bX_{1}$ then $h(x)$ is essentially at least as random with respect to the
output distribution $h^{*}\mu$ in $\bX_{0}$.
Decrease in randomness
can only be caused by complexity in the definition of the function $h$.
It is even easier to interpret the theorem when $\mu$ is defined over a product
space $\bX_{1}\times \bX_{2}$, and $h(x_{1},x_{2}) = x_{1}$ is the projection.
The theorem then says, informally, that if the pair $(x_{1},x_{2})$ is
random with respect to $\mu$ then $x_{1}$ is random with respect to the
marginal $\mu_{1} = h^{*}\mu$ of $\mu$.
This is a very natural requirement: why would the throwing-away of the
information about $x_{2}$ affect the plausibility of the
hypothesis that the outcome $x_{1}$ arose from the distribution 
$\mu_{1}$? 

In the general case of the theorem, concerning random transitions,
we cannot bound the randomness of each outcome uniformly.
The theorem asserts that the average nonrandomness, as measured by 
the universal test with respect to the output distribution, does not
increase.
In logarithmic notation:
$\lg_{x}^{y} 2^{\d_{\Lg^{*}\mu}(y)} \lea \d_{\mu}(x)$, 
or equivalently, 
$\int 2^{\d_{\Lg^{*}\mu}(y)} \lg_{x}(dy) \lea \d_{\mu}(x)$.

 \begin{corollary}
Let $\Lg$ be a computable probability kernel from $\bX_{1}$ to $\bX_{0}$.
There is a constant $c$ such that
for every $x\in\bX^{1}$, and integer $m > 0$ we have 
 \[
 \lg_{x}\setof{y : \d_{\Lg^{*}\mu}(y) > \d_{\mu}(x) + m + c} \le 2^{-m}.
 \]
 \end{corollary}

Thus, in a computable random transition,
the probability of an increase of randomness deficiency by
$m$ units (plus a constant $c$) is less than $2^{-m}$.
The constant $c$ comes from the description complexity 
of the transition $\Lg$.

A randomness conservation result related to 
Corollary~\ref{c.conservation-determ} was proved 
in~\cite{HertlingWeihrauchRand98}.
There, the measure over the space $\bX_{0}$ is not the output
measure of the transformation, but is assumed to
obey certain inequalities related to the transformation.

 \section{Tests and complexity}\label{s.complexity}

\subsection{Description complexity}

\subsubsection{Complexity, semimeasures, algorithmic entropy}
Let $X=\Sg^{*}$.
For $x \in \Sg^{*}$ for some finite alphabet $\Sg$,
let $H(x)$ denote the prefix-free description complexity of the finite
sequence $x$ as defined, for example, in~\cite{LiViBook97} (where it is
denoted by $K(x)$).
For completeness, we give its definition here.
Let $A : \{0,1\}^{*} \times \Sg^{*} \to \Sg^{*}$ be a computable
(possibly partial) function with the property that if $A(p_{1},y)$ and
$A(p_{2},y)$ are defined for two different strings $p_{1}, p_{2}$, then
$p_{1}$ is not the prefix of $p_{2}$.
Such a function is called a (prefix-free) \df{interpreter}.
We denote
 \[
   H^{A}(x \mid y) = \min_{A(p,y)=x} |p|.
 \]
One of the most important theorems of description complexity is the
following:

 \begin{proposition}[Invariance Theorem, see for
example~\protect\cite{LiViBook97}]
There is an optimal interpreter $T$ with the above property: with it, for
every interpreter $A$ there is a constant $c_{A}$ with
 \[
   H^{T}(x \mid y) \le H^{A}(x \mid y) + c_{A}.
 \]
 \end{proposition}

We fix an optimal interpreter $T$ and write 
$H(x \mid y) = H^{T}(x \mid y)$, calling it 
the conditional complexity of a string $x$ with respect to string $y$.
We denote $H(x) = H(x \mid \Lg)$.
Let 
 \[
   \m(x) = 2^{-H(x)}.
 \]
The function $\m(x)$ is lower semicomputable with
$\sum_{x} \m(x) \le 1$.
Let us call any real function $f(x) \ge 0$ over $\Sg^{*}$ with 
$\sum_{x} f(x) \le 1$ a \df{semimeasure}.
The following theorem, known as the Coding Theorem, is
an important tool.
  
 \begin{proposition}[Coding Theorem]\label{t.coding}
For every lower semicomputable semimeasure $f$ there is a constant $c>0$
with $\m(x) \ge c\cdot f(x)$.
 \end{proposition}

Because of this theorem, we will say that $\m(x)$ is a
\df{universal lower semicomputable semimeasure}.
It is possible to turn $\m(x)$ into a measure, by compactifying the
discrete space $\Sg^{*}$ into 
 \[
  \ol{\Sg^{*}}=\Sg^{*}\cup\{\infty\}
 \]
(as in part~\ref{i.compact.compactify} of Example~\ref{x.compact};
this process makes sense also for a constructive discrete space),
and setting $\m(\infty) = 1-\sum_{x\in\Sg^{*}} \m(x)$.
The extended measure $\m$ is not quite lower semicomputable since
the number $\mu(\ol{\Sg^{*}} \xcpt \{0\})$ is not necessarily
lower semicomputable.

 \begin{remark}
A measure $\mu$ is computable over $\ol{\Sg^{*}}$ if and only if the
function $x \mapsto \mu(x)$ is computable for $x \in \Sg^{*}$.
This property does not imply that the number
 \[
 1 - \mu(\infty) = \mu(\Sg^{*}) = \sum_{x\in\Sg^{*}} \mu(x)
 \]
 is computable.
 \end{remark}

Let us allow, for a moment, measures $\mu$ that are not probability
measures: they may not even be finite.
Metric and computability can be extended to this case
(see~\cite{Topsoe70}), the universal test
$\t_{\mu}(x)$ can also be generalized.
The Coding Theorem and other considerations suggest the introduction of
the following notation, for an arbitrary measure $\mu$:
 \begin{equation}\label{e.alg-ent}
   H_{\mu}(x) = -\d_{\mu}(x) = -\log\t_{\mu}(x).
 \end{equation}
Then, with $\#$ defined as the counting measure over the discrete set
$\Sg^{*}$ (that is, $\#(S) = |S|$), we have
 \[
   H(x) \eqa H_{\#}(x).
 \]
This allows viewing $H_{\mu}(x)$ as a generalization of description
complexity: we will call this quantity the \df{algorithmic entropy} of $x$
relative to the measure $\mu$.
Generalization to conditional complexity is done using
Remark~\ref{r.cond-test}.
A reformulation of the definition of tests says that $H_{\mu}(x)$
is minimal (within an additive constant) among the upper semicomputable
functions $(\mu, x) \mapsto f_{\mu}(x)$ with $\mu^{x} 2^{-f_{\mu}(x)} \le 1$.
The following identity is immediate from the definitions:
 \begin{equation}\label{e.Hmu-to-cond}
   H_{\mu}(x) = H_{\mu}(x \mid \mu).
 \end{equation}

\subsubsection{Computable measures and complexity}
It is known that for computable $\mu$, the test $\d_{\mu}(x)$ can be
expressed in terms of the description complexity of $x$
(we will prove these expressions below).
Assume that $\bX$ is the (discrete) space of all binary strings.
Then we have 
\begin{equation}\label{e.test-charac-fin-cpt}
   \d_{\mu}(x) = -\log \mu(x) - H(x) + O(H(\mu)).
\end{equation}
The meaning of this equation is the following.  
Due to maximality property of the semimeasure $\m$ following from
the Coding Theorem~\ref{t.coding} above, the expression
$-\log\mu(x)$ is an upper bound (within $O(H(\mu))$) of the
complexity $H(x)$, and nonrandomness of $x$ is measured by
the difference between the complexity and this upper bound.
See~\cite{ZvLe70} for a first formulation of this general upper bound
relation.
As a simple example, consider the uniform distribution $\mu$ over the set of
binary sequences of length $n$. 
Conditioning everything on $n$, we obtain 
 \[
  \d_{\mu}(x\mid n) \eqa n - H(x\mid n),
 \]
that is the more the description complexity $H(x\mid n)$ of a binary
sequence of length $n$ differs from its upper bound $n$ the less random is
$x$.

Assume that $\bX$ is the space of infinite binary sequences.
Then equation~\eqref{e.test-charac-fin-cpt} must be replaced with
\begin{equation}\label{e.test-charac-infin-cpt}
   \d_{\mu}(x) =
 \sup_{n}\Paren{-\log \mu(x^{\le n}) - H(x^{\le n})} + O(H(\mu)).
\end{equation}
For the coin-tossing distribution $\mu$, this characterization has first
been first proved by Schnorr, and published in~\cite{Chaitin75}.

 \begin{remark}\label{r.monot-compl}
It is possible to obtain similar natural characterizations of randomness,
using some other natural definitions of description complexity.
A universal semicomputable semimeasure
$\m_{\Og}$ over the set $\Og$ of infinite sequences was introduced, 
and a complexity $\KM(x) = -\log\m_{\Og}(x)$ defined in~\cite{ZvLe70}.
A so-called ``monotonic complexity'', $\Km(x)$ was introduced, using
Turing machines with one-way input and output,
in~\cite{LevinRand73}, and a closely
related quantity called ``process complexity'' was introduced
in~\cite{Schnorr73}.
These quantities can also be used in a characterization of randomness
similar to~\eqref{e.test-charac-fin-cpt}.
The nontrivial fact that the complexities $\KM$ and $\Km$ differ by an
unbounded amount was shown in~\cite{GacsRel83}.
 \end{remark}

For noncomputable measures, we cannot replace $O(H(\mu))$ in
these relations with anything finite, as shown in
the following example.
Therefore however attractive and simple, 
$\exp(-\log \mu(x) - H(x))$ is not a universal uniform test of randomness. 

 \begin{proposition}\label{p.test-charac-counterexample}
There is a measure $\mu$ over the discrete space $\bX$ of binary strings
such that for each $n$, there is an $x$ with
$\d_{\mu}(x) = n - H(n)$ and $-\log \mu(x) - H(x) \lea 0$.
 \end{proposition}
 \begin{proof}
Let us treat the domain of our measure $\mu$ as a set of pairs $(x,y)$.
Let $x_{n} = 0^{n}$, for $n=1,2,\dotsc$.
For each $n$, let $y_{n}$ be some binary string of length $n$
with the property $H(x_{n}, y_{n}) > n$.
Let $\mu(x_{n},y_{n})=2^{-n}$.
Then $- \log \mu(x_{n},y_{n}) - H(x_{n},y_{n}) \le n - n = 0$.
On the other hand, let $t_{\mu}(x,y)$ be the test nonzero only on strings
$x$ of the form $x_{n}$:
 \[
 t_{\mu}(x_{n}, y) = \frac{\m(n)}{\sum_{z \in \cB^{n}} \mu(x_{n}, z)}.
 \]
The form of the definition ensures semicomputability and we also have
 \[
 \sum_{x,y} \mu(x,y) t_{\mu}(x,y) \le \sum_{n} \m(n) < 1,
 \]
therefore $t_{\mu}$ is indeed a test.
Hence $\t_{\mu}(x,y) \gem t_{\mu}(x,y)$.
Taking logarithms, $\d_{\mu}(x_{n}, y_{n}) \gea n - H(n)$.
 \end{proof}

 The same example implies that it is also not an option, even over discrete
 sets, to replace the definition of uniform tests with the \emph{ad hoc}
 formula $\exp(-\log\mu(x) - H(x))$:

 \begin{proposition}
The test defined as $f_{\mu}(x) = \exp(-\log\mu(x) - H(x))$ 
over discrete spaces $\bX$ does
not obey the conservation of randomness.
 \end{proposition}
 \begin{proof}
Let us use the example of Proposition~\ref{p.test-charac-counterexample}.
Consider the function $\pi : (x,y) \mapsto x$.
The image of the measure $\mu$ under the projection is 
$(\pi\mu)(x) = \sum_{y} \mu(x,y)$.
Thus, $(\pi\mu)(x_{n}) = \mu(x_{n},y_{n}) = 2^{-n}$.
We have seen $\log f_{\mu}(x_{n},y_{n}) \le 0$.
On the other hand,
 \[ 
 \log f_{\pi\mu} (\pi(x_{n},y_{n})) = -\log(\pi\mu)(x_{n}) - H(x_{n})
 \eqa n - H(n). 
 \]
Thus, the projection $\pi$ takes a random pair $(x_{n},y_{n})$ into
an object $x_{n}$ that is very nonrandom (when randomness is measured using
the tests $f_{\mu}$).
 \end{proof}
In the example, we have the abnormal situation that a pair is random but
one of its elements is nonrandom.
Therefore even if we would not insist on universality, the test 
$\exp(-\log\mu(x) - H(x))$ is unsatisfactory.

Looking into the reasons of the nonconservation in the example,
we will notice that it could only have happened because the 
test $f_{\mu}$ is too special.
The fact that $-\log (\pi\mu)(x_{n}) - H(x_{n})$ is large should show that 
the pair $(x_{n},y_{n})$ can be enclosed into the ``simple'' set
$\{x_{n}\} \times \bY$ of small probability; unfortunately, 
this observation does not reflect on $-\log\mu(x,y) - H(x,y)$ 
when the measure $\mu$ is non-computable (it does for computable $\mu$).
 
\subsubsection{Expressing the uniform test in terms of complexity}
It is a natural idea to modify equation~\eqref{e.test-charac-fin-cpt}
in such a way that the complexity $H(x)$ is replaced with $H(x \mid \mu)$.
However, this expression must be understood properly.
The measure $\mu$
(especially, when it is not computable) cannot be described by
a finite string; on the other hand, it can be described by infinite strings
in many different ways.
Clearly, irrelevant information in these infinite strings should be
ignored.
The notion of representation in computable analysis (see 
Subsection~\ref{ss.notation-repr}) will solve the problem.
An interpreter function should have the property that its output depends
only on $\mu$ and not on the sequence representing it.
Recall the topological space $\bM$
of computable measures over our space $\bX$.
An interpreter $A : \{0,1\}^{*} \times \bM \to \Sg^{*}$ is 
a computable function that is prefix-free in its first argument.
The complexity 
 \[
  H(x \mid \mu)
 \]
can now be defined in terms of such
interpreters, noting that the Invariance Theorem holds as before.
To define this complexity in terms of representations,
let $\gm_{\bM}$ be our chosen representation for the space $\bM$
(thus, each measure $\mu$ is represented via all of its
Cauchy sequences in the Prokhorov distance).
Then we can say that $A$ is an interpreter if it is
$(\id, \gm_{\bM}, \id)$-computable, that is a certain computable function
$B : \{0,1\}^{*} \times \Sg^{\og} \to \Sg^{*}$ realizes $A$
for every $p\in\{0,1\}^{*}$,
and for every sequence $z$ that is a $\gm_{\bM}$-name of a measure $\mu$, we
have $B(p, z) = A(p, \mu)$.

 \begin{remark}
The notion of oracle computation and reducibility in
the new sense (where the result is 
required to be independent of which representation of an
object is used) may be worth investigating in other settings as well.
 \end{remark}

Let us mention the following easy fact:

 \begin{proposition}\label{p.H/mu-computable}
If $\mu$ is a computable measure then $H(x \mid \mu) \eqa H(x)$.
The constant in $\eqa$ depends on the description complexity of $\mu$.
 \end{proposition}

 \begin{theorem}\label{t.test-charac-discr}
If $\bX$ is the discrete space $\Sg^{*}$ then we have
 \begin{equation}\label{e.test-charac-discr}
   \d_{\mu}(x) \eqa -\log\mu(x) - H(x \mid \mu).
 \end{equation}
 \end{theorem}

Note that in terms of the algorithmic entropy notation introduced 
in~\eqref{e.alg-ent}, this theorem can be expressed as
 \begin{equation}\label{e.alg-entr-charac}
   H_{\mu}(x) \eqa H(x \mid \mu) + \log\mu(x).
 \end{equation}
 \begin{proof}
In exponential notation, equation~\eqref{e.test-charac-discr} can be
written as $\t_{\mu}(x) \eqm \m(x \mid \mu)/\mu(x)$.
Let us prove $\gem$ first.
We will show that the right-hand side of this inequality is a test, and
hence $\lem \t_{\mu}(x)$.
However, 
the right-hand side is clearly lower semicomputable in $(x, \mu)$ and
when we ``integrate'' it (multiply it by $\mu(x)$ and sum it), its sum is
$\le 1$; thus, it is a test.

Let us prove $\lem$ now.
The expression $\t_{\mu}(x)\mu(x)$ is clearly lower semicomputable in
$(x,\mu)$, and its sum is $\le 1$.
Hence, it is $\lea \m(x \mid \mu)$.
 \end{proof}

 \begin{remark}
As mentioned earlier, our theory generalizes to measures that are not
probability measures.
In this case, equation~\eqref{e.alg-entr-charac} has interesting relations to
the quantity called ``physical entropy'' by Zurek in~\cite{ZurekPhR89};
it justifies calling $H_{\mu}(x)$
``fine-grained algorithmic Boltzmann entropy'' by this author 
in~\cite{GacsBoltzmann94}.
 \end{remark}

For non-discrete spaces, unfortunately, we can only provide less
intuitive expressions.

 \begin{proposition}\label{t.test-charac}
let $\bX=(X, d, D, \ag)$ be a complete
computable metric space, and let $\cE$ be the enumerated set of bounded
Lip\-schitz functions introduced in~\eqref{e.bd-Lip-seq}, 
but for the space $\bM(\bX) \times \bX$.
The uniform test of randomness $\t_{\mu}(x)$ can be expressed as
 \begin{equation}\label{e.test-charac}
  \t_{\mu}(x) \eqm 
  \sum_{f \in \cE}f(\mu,x)\frac{\m(f \mid \mu)}{\mu^{y} f(\mu, y)}.
 \end{equation}
 \end{proposition}
 \begin{proof}
For $\gem$,
we will show that the right-hand side of the inequality is a test, and
hence $\lem \t_{\mu}(x)$.
For simplicity, we skip the notation for the enumeration of $\cE$ and
treat each element $f$ as its own name.
Each term of the sum is clearly lower semicomputable in
$(f, x, \mu)$, hence the sum is lower semicomputable in $(x, \mu)$.
It remains to show that the $\mu$-integral of the sum is $\le 1$.
But, the $\mu$-integral of the generic term is $\le \m(f \mid \mu)$, and
the sum of these terms is $\le 1$ by the definition of the function
$\m(\cdot \mid \cdot)$.
Thus, the sum is a test.

For $\lem$, note that $(\mu,x) \mapsto \t_{\mu}(x)$, 
as a lower semicomputable function, is the supremum of functions in $\cE$.
Denoting their differences by $f_{i}(\mu,x)$,
we have $\t_{\mu}(x) = \sum_{i} f_{i}(\mu,x)$.
The test property implies $\sum_{i} \mu^{x} f_{i}(\mu,x) \le 1$.
Since the function $(\mu,i) \mapsto \mu^{x} f_{i}(\mu,x)$ is lower
semicomputable, this implies $\mu^{x} f_{i}(\mu,x) \lem \m(i \mid \mu)$, and
hence 
 \[
 f_{i}(\mu,x) \lem f_{i}(\mu,x) \frac{\m(i \mid \mu)}{\mu^{x} f_{i}(\mu,x)}.
 \]
It is easy to see that for each $f\in\cE$ we have
 \[
   \sum_{i : f_{i} = f} \m(i \mid \mu) \le \mu(f \mid \mu),
 \]
which leads to~\eqref{e.test-charac}.
 \end{proof}

 \begin{remark}\label{r.test-charac-lb}
If we only want the $\gem$ part of the result, then $\cE$ can be
replaced with any enumerated computable sequence of bounded computable
functions.
 \end{remark}

  \subsection{Infinite sequences}

In this section, we get a nicer characterization of randomness tests in
terms of complexity, in special cases.
Let $\cM_{R}(X)$ be the set of measures $\mu$ with $\mu(X)=R$.

 \begin{theorem}\label{t.defic-charac-cpt-seqs}
Let $\bX=\bbN^{\og}$ be the set of infinite sequences of natural numbers,
with the product topology.
For all computable measures $\mu\in\cM_{R}(X)$, 
for the deficiency of randomness $\d_{\mu}(x)$, we have
 \begin{equation}\label{e.defic-charac-seq}
 \d_{\mu}(x) \eqa \sup_{n}\Paren{-\log \mu(x^{\le n}) - H(x^{\le n})}.
 \end{equation}
Here, the constant in $\eqa$ depends on the computable measure $\mu$.
 \end{theorem}

We will be able to prove the $\gea$ part of the statement in a more general
space, and without assuming computability.
Assume that a separating sequence $b_{1},b_{2},\dots$ is given as defined
in Subsection~\ref{ss.cells}, along with the set $X^{0}$.
For each $x \in X^{0}$, the binary sequence $x_{1},x_{2},\dots$ 
has been defined.
Let
 \begin{align*}
  \ol\mu(\Gg_{s}) &= R  - \sum\setof{\mu(\Gg_{s'}) : l(s)=l(s'),\;s'\ne s}.
 \end{align*}
Then $(s,\mu)\mapsto \mu(\Gg_{s})$ is lower semicomputable, and 
$(s,\mu)\mapsto \ol\mu(\Gg_{s})$ is upper semicomputable.
And, every time that the functions $b_{i}(x)$ form a regular partition for
$\mu$, we have $\ol\mu(\Gg_{s})=\mu(\Gg_{s})$ for all $s$.
Let $\cM_{R}^{0}(X)$ be the set of those measures $\mu$
in $\cM_{R}(X)$ for which $\mu(X \xcpt X^{0})=0$.

 \begin{theorem}\label{t.defic-charac-compact}
Suppose that the space $\bX$ is compact.
Then for all computable measures $\mu\in\cM_{R}^{0}(\bX)$, 
for the deficiency of randomness $\d_{\mu}(x)$, 
the characterization~\eqref{e.defic-charac-seq} holds.
 \end{theorem}

For arbitrary measures and spaces, we can say a little less:

 \begin{proposition}\label{p.test-charac-seq-lb}
For all measures $\mu\in\cM_{R}(X)$, 
for the deficiency of randomness $\d_{\mu}(x)$, we have
 \begin{equation}\label{e.defic-ineq-seq}
 \d_{\mu}(x) \gea \sup_{n}\Paren{-\log \ol\mu(x^{\le n}) - H(x^{\le n} \mid \mu)}.
 \end{equation}
 \end{proposition}
 \begin{proof}
  Consider the function
 \[
 f_{\mu}(x) = \sum_{s} 1_{\Gg_{s}}(x) \frac{\m(s \mid \mu)}{\ol\mu(\Gg_{s})}
  = \sum_{n} \frac{\m(x^{\le n} \mid \mu)}{\ol\mu(x^{\le n})}
  \ge \sup_{n} \frac{\m(x^{\le n} \mid \mu)}{\ol\mu(x^{\le n})}.
 \]
The function $(\mu,x) \mapsto f_{\mu}(x)$ 
is clearly lower semicomputable and satisfies
$\mu^{x} f_{\mu}(x) \le 1$, and hence 
 \[
  \d_{\mu}(x) \gea \log f(x) \gea 
   \sup_{n}\Paren{-\log\ol\mu(x^{\le n}) - H(x^{\le n} \mid \mu)}.
 \]
 \end{proof}

 \begin{proof}[Proof of Theorem~\protect\ref{t.defic-charac-cpt-seqs}]
For binary sequences instead of sequences of natural numbers, the part
$\gea$ of the inequality follows directly from
Proposition~\ref{p.test-charac-seq-lb}: indeed, look at
Examples~\ref{x.cells}.
For sequences of natural numbers, the proof is completely analogous.

The proof of $\lea$ reproduces the proof of Theorem 5.2
of~\cite{GacsExact80}.
The computability of $\mu$ implies that $t(x)=\t_{\mu}(x)$ is lower
semicomputable.
Let us first replace $t(x)$ with a rougher version:
 \[
   t'(x) = \max \setof{2^{n} : 2^{n} < \t_{\mu}(x)}.
 \]
Then $t'(x) \eqm t(x)$, and it takes only values of the form $2^{n}$.
It is also lower semicomputable.
Let us abbreviate:
 \[
   1_{y}(x) = 1_{x\bbN^{\og}}(x),\quad \mu(y) = \mu(y\bbN^{\og}).
 \]
For every lower semicomputable function $f$ over $\bbN^{\og}$, 
there are computable sequences $y_{i}\in\bbN^{*}$ and
$r_{i}\in\bbQ$ with $f(x) = \sup_{i} r_{i} 1_{y_{i}}(x)$, with the
additional property that if $i<j$ and $1_{y_{i}}(x)=1_{y_{j}}(x)=1$ then
$r_{i}<r_{j}$.
Since $t'(x)$ only takes values of the form $2^{n}$, there are computable
sequences $y_{i} \in \bbB^{*}$ and $k_{i} \in \bbN$ with
 \[
 t'(x) = \sup_{i}\; 2^{k_{i}} 1_{y_{i}}(x) 
  \eqm \sum_{i} 2^{k_{i}} 1_{y_{i}}(x),
 \]
with the property that if $i<j$ and $1_{y_{i}}(x)=1_{y_{j}}(x)=1$ then
$k_{i}<k_{j}$.
The equality $\eqm$ follows easily from the fact that for any finite
sequence 
$n_{1}<n_{2}<\dots$, $\sum_{j} 2^{n_{j}} \le 2 \max_{j} 2^{n_{j}}$.
Since $\mu t' \lem 1$, we have $\sum_{i} 2^{k_{i}}\mu(y_{i}) \lem 1$.
Since the function $i \to 2^{k_{i}}\mu(y_{i})$ is computable, this implies
$2^{k_{i}}\mu(y_{i}) \lem \m(i)$, $2^{k_{i}} \lem \m(i)/\m(y_{i})$.
Thus,
 \[
   t(x) \lem \sup_{i}\; 1_{y_{i}}(x)\frac{\m(i)}{\mu(y_{i})}.
 \]
For $y \in \bbN^{*}$ we certainly have 
$H(y) \lea \inf_{y = y_{i}} H(i)$, which implies
$\sup_{y = y_{i}} \m(i) \le \m(y)$.
It follows that
 \[
   t(x) \lem \sup_{y \in \bbB^{*}}\; 1_{y}(x)\frac{\m(y)}{\mu(y)}
  = \sup_{n} \frac{\m(x^{\le n})}{\mu(x^{\le n})}.
 \]
Taking logarithms, we obtain the $\lea$ part of the theorem.
 \end{proof}

 \begin{proof}[Proof of Theorem~\protect\ref{t.defic-charac-compact}]
The proof of part $\gea$ of the inequality follows directly from
Proposition~\ref{p.test-charac-seq-lb}, just as in the proof of 
Theorem~\ref{t.defic-charac-cpt-seqs}.

The proof of $\lea$ is also similar to the proof of that theorem.
The only part that needs to be reproved is the statement that
for every lower semicomputable function $f$ over $X$,
there are computable sequences $y_{i}\in\bbB^{*}$ and
$q_{i}\in\bbQ$ with $f(x) = \sup_{i} q_{i} 1_{y_{i}}(x)$.
This follows now, since according to
Proposition~\ref{p.compact-cell-basis}, the cells $\Gg_{y}$ form a basis of
the space $\bX$.
 \end{proof}

\section{Neutral measure}\label{s.neutral}

Let $\t_{\mu}(x)$ be our universal uniform randomness test.
We call a measure $M$ \df{neutral} if
$\t_{M}(x) \le 1$ for all $x$.
If $M$ is neutral then no experimental outcome $x$ could 
refute the theory (hypothesis, model)
that $M$ is the underlying measure to our experiments.
It can be used as ``apriori probability'', in a Bayesian approach to
statistics.
Levin's theorem says the following:

 \begin{theorem}\label{t.neutral-meas}
If the space $\bX$ is compact then there is a neutral measure over $\bX$.
 \end{theorem}
 The proof relies on a nontrivial combinatorial fact,
Sperner's Lemma, which also underlies the proof of the Brouwer fixpoint
theorem.
Here is a version of Sperner's Lemma, spelled out in continuous form:

 \begin{proposition}[see for
     example~\protect\cite{SpanierAlgTop71}]\label{p.Sperner} 
Let $p_{1},\dots,p_{k}$ be points of some finite-dimensional space
$\bbR^{n}$.
Suppose that there are closed sets $F_{1},\dots,F_{k}$
with the property that for every subset $1 \le i_{1} < \dots < i_{j} \le k$
of the indices, the simplex $S(p_{i_{1}},\dots,p_{i_{j}})$ spanned by
$p_{i_{1}},\dots,p_{i_{j}}$ is covered by
the union $F_{i_{1}} \cup \dots \cup F_{i_{j}}$.
Then the intersection $\bigcap_{i} F_{i}$ of all these sets is not empty.
 \end{proposition}

The following lemma will also be needed.

 \begin{lemma}\label{l.concentr}
For every closed set $A \sbs \bX$ and measure $\mu$, if $\mu(A)=1$ then
there is a point $x\in A$ with $\t_{\mu}(x) \le 1$.
 \end{lemma}
 \begin{proof}
This follows easily from 
$\mu\, t_{\mu} = \mu^{x} 1_{A}(x)t_{\mu}(x) \le 1$.
 \end{proof} 

 \begin{proof}[Proof of Theorem~\protect\ref{t.neutral-meas}]
For every point $x \in \bX$, let $F_{x}$ be the set of measures for which
$\t_{\mu}(x) \le 1$.
If we show that for every finite set of points $x_{1},\dots,x_{k}$, we
have 
 \begin{equation}\label{e.finite-inters}
  F_{x_{1}}\cap\dots\cap F_{x_{k}} \ne \emptyset,
 \end{equation}
then we will be done.
Indeed, according to Proposition~\ref{p.measures-compact}, the compactness
of $\bX$ implies the compactness of the space $\cM(\bX)$ of measures.
Therefore if every finite subset of the family $\setof{F_{x} : x \in \bX}$
of closed sets has a nonempty intersection, then the whole family has a
nonempty intersection: this intersection consists of the neutral
measures.

To show~\eqref{e.finite-inters}, let $S(x_{1},\dots,x_{k})$ be the set of
probability measures concentrated on $x_{1},\dots,x_{k}$.
Lemma~\ref{l.concentr} implies that each such measure belongs to one of
the sets $F_{x_{i}}$.
Hence $S(x_{1},\dots,x_{k}) \sbs F_{x_{1}} \cup \dots \cup F_{x_{k}}$,
and the same holds for every subset of the indices $\{1,\dots,k\}$.
Sperner's Lemma~\ref{p.Sperner} implies 
$F_{x_{1}} \cap \dots \cap F_{x_{k}} \ne\emptyset$.
 \end{proof}

When the space is not compact, there are generally no neutral probability
measures, as shown by the following example.

 \begin{proposition}\label{t.no-neutral}
Over the discrete space $\bX = \bbN$ of natural numbers,
there is no neutral measure.
 \end{proposition}
 \begin{proof}
It is sufficient to 
construct a randomness test $t_{\mu}(x)$ with the property that for
every measure $\mu$, we have $\sup_{x} t_{\mu}(x) = \infty$.
Let 
 \begin{equation}\label{e.no-neutral}
   t_{\mu}(x) = \sup\setof{ k \in \bbN: \sum_{y<x}\mu(y) > 1-2^{-k}}.
 \end{equation}
By its construction, this is a lower semicomputable function with
$\sup_{x} t_{\mu}(x) = \infty$.
It is a test if $\sum_{x}\mu(x)t_{\mu}(x) \le 1$.
We have
 \[
  \sum_{x} \mu(x) t_{\mu}(x) 
= \sum_{k>0} \sum_{t_{\mu}(x) \ge k} \mu(x)
< \sum_{k>0} 2^{-k} \le 1.
 \]
 \end{proof}

Using a similar construction over the space $\bbN^{\og}$ of infinite
sequences of natural numbers, we could show that 
for every measure $\mu$ there is a sequence $x$ with $\t_{\mu}(x)=\infty$.

Proposition~\ref{t.no-neutral} is a little misleading, since as a locally
compact set, $\bbN$ can be compactified into $\ol\bbN = \bbN \cup \{\infty\}$
(as in Part~\ref{i.compact.compactify} of Example~\ref{x.compact}).
Theorem~\ref{t.neutral-meas} implies that there is a neutral probability
measure $M$ over the compactified space $\ol\bbN$.
Its restriction to $\bbN$ is, of course, not a probability measure, since it
satisfies only $\sum_{x < \infty} M(x) \le 1$.
We called these functions \df{semimeasures}.

 \begin{remark}\label{r.compactify}\
 \begin{enumerate}[\upshape 1.]
   \item
 It is easy to see that 
Theorem~\ref{t.test-charac-discr} characterizing randomness in terms of
complexity holds also for the space $\ol\bbN$.

   \item
The topological space of semimeasures
over $\bbN$ is not compact, and there is no neutral one among them.
Its topology is not the same as what we get when we restrict the topology
of probability measures over $\ol\bbN$ to $\bbN$.
The difference is that over $\bbN$, for example the set of measures
$\setof{\mu : \mu(\bbN) \ge 1/2}$ is closed, since $\bbN$ (as the whole space)
is a closed set.
But over $\ol\bbN$, this set is not closed.
 \end{enumerate}
 \end{remark}

Neutral measures are not too simple, even over $\ol\bbN$, as the following
theorem shows.

 \begin{theorem}\label{t.no-upper-semi-neutral}
There is no neutral measure over $\ol\bbN$ that is upper
semicomputable over $\bbN$ or lower semicomputable over $\bbN$.
 \end{theorem}
 \begin{proof}
Let us assume that $\nu$ is a measure that is upper semicomputable over
$\bbN$.
Then the set
 \[
   \setof{(x,r) : x \in\bbN,\; r\in\bbQ,\; \nu(x) < r}
 \]
is recursively enumerable: let $(x_{i},r_{i})$ be a particular
enumeration. 
For each $n$, let $i(n)$ be the first $i$ with $r_{i} < 2^{-n}$, and let
$y_{n} = x_{i(n)}$.
Then $\nu(y_{n}) < 2^{-n}$, and at the same time $H(y_{n}) \lea H(n)$.
As mentioned, in Remark~\ref{r.compactify}, 
Theorem~\ref{t.test-charac-discr} characterizing randomness in terms of
complexity holds also for the space $\ol\bbN$.
Thus, 
 \[
  \d_{\nu}(y_{n}) \eqa -\log\nu(y_{n}) - H(y_{n} \mid \nu) \gea n - H(n).
 \]
Suppose now that $\nu$ is lower semicomputable over $\bbN$.
The proof for this case is longer.
We know that $\nu$ is the monotonic limit of a recursive sequence
$i\mapsto \nu_{i}(x)$ of recursive semimeasures with rational values
$\nu_{i}(x)$.
For every $k=0,\dots,2^{n}-2$, let 
 \begin{align*}
  V_{n,k} &= \setof{\mu \in \cM(\ol\bbN) : 
k\cdot 2^{-n} < \mu(\{0,\dots,2^{n}-1\}) < (k+2)\cdot 2^{-n}},
\\       \J &= \setof{(n,k): k\cdot 2^{-n} < \nu(\{0,\dots,2^{n}-1\})}.
 \end{align*}
The set $\J$ is recursively enumerable.
Let us define the functions $j:\J\to\bbN$ and $x:\J\to\{0,\dots,2^{n}-1\}$
as follows: $j(n,k)$ is the smallest $i$ with 
$\nu_{i}(\{0,\dots,2^{n}-1\}) > k\cdot 2^{-n}$, and
 \[
   x_{n,k} = \min\setof{y < 2^{n}: \nu_{j(n,k)}(y) < 2^{-n+1}}.
 \]
Let us define the function $f_{\mu}(x,n,k)$ as follows.
We set $f_{\mu}(x,n,k)=2^{n-2}$ if the following conditions hold:
 \begin{enumerate}[(a)]
  \item\label{i.no-upper-semi-neutral.mu-global} $\mu \in V_{n,k}$;
  \item\label{i.no-upper-semi-neutral.mu-upper} $\mu(x) < 2^{-n+2}$;
  \item\label{i.no-upper-semi-neutral.unique}
 $(n,k) \in \J$ and $x=x_{n,k}$.
 \end{enumerate}
Otherwise, $f_{\mu}(x,n,k)=0$.
Clearly, the function $(\mu,x,n,k) \mapsto f_{\mu}(x,n,k)$ is lower
semicomputable.
Condition~\eqref{i.no-upper-semi-neutral.mu-upper} implies
 \begin{equation}\label{e.no-upper-semi-neutral.n-k-test}
 \sum_{y} \mu(y) f_{\mu}(y,n,k) \le
    \mu(x_{n,k})f_{\mu}(x_{n,k},n,k) < 2^{-n+2}\cdot 2^{n-2} = 1.
 \end{equation}
Let us show that $\nu \in V_{n,k}$ implies
 \begin{equation}\label{e.found-bad}
 f_{\nu}(x_{n,k},n,k) = 2^{n-2}.
 \end{equation}
Consider $x=x_{n,k}$.
Conditions~\eqref{i.no-upper-semi-neutral.mu-global} 
and~\eqref{i.no-upper-semi-neutral.unique} are satisfied by definition.
Let us show that condition~\eqref{i.no-upper-semi-neutral.mu-upper} is also 
satisfied.
Let $j=j(n,k)$.
By definition, we have $\nu_{j}(x) < 2^{-n+1}$.
Since by definition $\nu_{j}\in V_{n,k}$ and $\nu_{j} \le \nu \in V_{n,k}$,
we have 
 \[
  \nu(x) \le \nu_{j}(x) + 2^{-n+1} < 2^{-n+1} + 2^{-n+1} = 2^{-n+2}.
 \]
Since all three conditions~\eqref{i.no-upper-semi-neutral.mu-global},
\eqref{i.no-upper-semi-neutral.mu-upper}
and~\eqref{i.no-upper-semi-neutral.unique} are satisfied, we have 
shown~\eqref{e.found-bad}.
Now we define
 \[
   g_{\mu}(x) = \sum_{n\ge 2}\frac{1}{n(n+1)}\sum_{k}f_{\mu}(x,n,k).
 \]
Let us prove that $g_{\mu}(x)$ is a uniform test.
It is lower semicomputable by definition, so we only need to prove
$\sum_{x} \mu(x) f_{\mu}(x) \le 1$.
For this, let $I_{n,\mu} = \setof{k: \mu\in V_{n,k}}$.
Clearly by definition, $|I_{n,\mu}|\le 2$.
We have, using this last fact and the test 
property~\eqref{e.no-upper-semi-neutral.n-k-test}:
 \[
  \sum_{x} \mu(x) g_{\mu}(x) = 
  \sum_{n\ge 2}\frac{1}{n(n+1)}
  \sum_{k\in I_{n,\mu}} \sum_{x}\mu(x) f_{\mu}(x,n,k)
       \le  \sum_{n\ge 2}\frac{1}{n(n+1)}\cdot 2 \le 1.
 \]
Thus, $g_{\mu}(x)$ is a uniform test.
If $\nu\in V_{n,k}$ then we have
 \[
 \t_{\nu}(x_{n,k}) 
\gem g_{\nu}(x_{n,k}) \ge \frac{1}{n(n+1)}f_{\mu}(x_{n,k},n,k) \ge 
  \frac{2^{n-2}}{n(n+1)}.
 \]
Hence $\nu$ is not neutral.
 \end{proof}

 \begin{remark}
In~\cite{LevinUnif76} and~\cite{LevinRandCons84},
Levin imposed extra conditions on tests which allow to find a lower
semicomputable neutral semimeasure.
A typical (doubtless reasonable)
consequence of these conditions would be that if outcome $x$ is
random with respect to measures $\mu$ and $\nu$ then it is also random with
respect to $(\mu+\nu)/2$.
 \end{remark}

 \begin{remark}
The universal lower semicomputable
semimeasure $\m(x)$ has a certain property similar to neutrality.
According to Theorem~\ref{t.test-charac-discr}, 
for every computable measure $\mu$ we have
$\d_{\mu}(x) \eqa -\log\mu(x) - H(x)$
(where the constant in $\eqa$ depends on $\mu$).
So, for computable measures, the expression
 \begin{equation}\label{e.ol-d}
  \ol\d_{\mu}(x) = -\log\mu(x) - H(x)
 \end{equation}
can serve as a reasonable deficiency of randomness.
(We will also use the test $\ol\t = 2^{\ol\d}$.)
If we substitute $\m$ for $\mu$ in $\ol\d_{\mu}(x)$, we get 0.
This substitution is not justified, of course.
The fact that $\m$ is not a probability 
measure can be helped, at least over $\bbN$, using compactification as
above, and extending the notion of randomness tests. 
But the test $\ol\d_{\mu}$ can replace $\d_{\mu}$
only for computable $\mu$, while $\m$ is not computable.
Anyway, this is the sense in which all outcomes might
be considered random with respect to $\m$, and the heuristic
sense in which $\m$ may still be considered ``neutral''.
 \end{remark}

 \begin{remark}
Solomonoff proposed the use of a universal lower semicomputable semimeasure
(actually, a closely related structure) for inductive inference 
in~\cite{Solomonoff64I}.
He proved in~\cite{Solomonoff78} that sequences emitted by any computable
probability distribution can be predicted well by his scheme.
It may be interesting to see whether the same prediction scheme has
stronger properties when used with the truly neutral measure $M$ of the
present paper.
 \end{remark}

\section{Relative entropy}\label{s.rel-entr}

Some properties of description complexity make it a good
expression of the idea of individual information content.

\subsection{Entropy}
The entropy of a discrete probability distribution $\mu$ is defined as
 \[
   \cH(\mu) = - \sum_{x} \mu(x) \log \mu(x).
 \]
To generalize entropy to continuous distributions the
\df{relative entropy} is defined as follows.
Let $\mu,\nu$ be two measures, where $\mu$ is taken (typically, but not
always), to be a probability measure, and $\nu$ another measure, that can
also be a probability measure but is most frequently not.
We define the \df{relative entropy} $\cH_{\nu}(\mu)$ as follows.
If $\mu$ is not absolutely continuous with respect to $\nu$ then 
$\cH_{\nu}(\mu) = -\infty$.
Otherwise, writing 
 \[
  \frac{d\mu}{d\nu} = \frac{\mu(dx)}{\nu(dx)} =: f(x)
 \] 
for the (Radon-Nikodym) derivative 
(density) of $\mu$ with respect to $\nu$, we define
 \[
   \cH_{\nu}(\mu) = - \int \log\frac{d\mu}{d\nu} d\mu 
     = - \mu^{x} \log\frac{\mu(dx)}{\nu(dx)} = -\nu^{x} f(x) \log f(x).
 \]
Thus, $\cH(\mu) = \cH_{\#}(\mu)$ is a special case.

 \begin{example}
 Let $f(x)$ be a probability density function for the distribution
$\mu$ over the real line, and let $\lg$ be the Lebesgue measure there.
Then
 \[
   \cH_{\lg}(\mu) = -\int f(x) \log f(x) d x.
 \]
 \end{example}

In information theory and statistics, when both $\mu$ and $\nu$ are
probability measures, then $-\cH_{\nu}(\mu)$ is also denoted
$D(\mu \parallel \nu)$, and called (after Kullback) the
information divergence of the two measures.
It is frequently used in the role of a distance between $\mu$ and $\nu$.
It is not symmetric, but can be shown to obey the triangle inequality, and
to be nonnegative.
Let us prove the latter property: in our terms, it says that relative
entropy is nonpositive when both $\mu$ and $\nu$ are probability measures.

 \begin{proposition}\label{p.Kullback-pos}
Over a space $\bX$, we have
 \begin{equation}\label{e.Kullback-pos}
   \cH_{\nu}(\mu) \le -\mu(X) \log\frac{\mu(X)}{\nu(X)}.
 \end{equation}
 In particular, if $\mu(X) \ge \nu(X)$ then $\cH_{\nu}(\mu) \le 0$.
 \end{proposition}
 \begin{proof}
The inequality $- a \ln a \le -a\ln b + b-a$
expresses the concavity of the logarithm function.
Substituting $a = f(x)$ and $b = \mu(X)/\nu(X)$ 
and integrating by $\nu$:
 \[
 (\ln 2) \cH_{\nu}(\mu) = 
 -\nu^{x} f(x) \ln f(x) \le -\mu(X) \ln\frac{\mu(X)}{\nu(X)}
  + \frac{\mu(X)}{\nu(X)} \nu(X) - \mu(X)
 = -\mu(X) \ln\frac{\mu(X)}{\nu(X)},
 \]
giving~\eqref{e.Kullback-pos}.
 \end{proof}

The following theorem generalizes an earlier known theorem stating that
over a discrete space, for a computable measure,
entropy is within an additive constant the same as ``average complexity'':
$\cH(\mu) \eqa \mu^{x} H(x)$.

 \begin{theorem}
Let $\mu$ be a probability measure.
Then we have
 \begin{equation}\label{e.entropy-less-avg-algentr}
  \cH_{\nu}(\mu) \le \mu^{x} H_{\nu}(x \mid \mu).
 \end{equation}
If $X$ is a discrete space then the following estimate also holds:
  \begin{equation}\label{e.entropy-gea-avg-algentr}
  \cH_{\nu}(\mu) \gea \mu^{x} H_{\nu}(x \mid \mu).
  \end{equation}
 \end{theorem}
 \begin{proof} 
Let
$\dg$ be the measure with density $\t_{\nu}(x \mid \mu)$ with respect to
$\nu$: $\t_{\nu}(x \mid \mu) = \frac{\dg(dx)}{\nu(dx)}$.
Then $\dg(X) \le 1$.
It is easy to see from the maximality property of $\t_{\nu}(x \mid \mu)$
that $\t_{\nu}(x \mid \mu) > 0$, therefore according to
Proposition~\ref{p.density-props}, we have
$\frac{\nu(dx)}{\dg(dx)} = \Paren{\frac{\dg(dx)}{\nu(dx)}}^{-1}$.
Using Proposition~\ref{p.density-props} and~\ref{p.Kullback-pos}: 
 \begin{align*}
     \cH_{\nu}(\mu)     &= - \mu^{x} \log\frac{\mu(dx)}{\nu(dx)},
\\   - \mu^{x} H_{\nu}(x \mid \mu) &=  \mu^{x} \log \frac{\dg(dx)}{\nu(dx)}
              = - \mu^{x} \log \frac{\nu(dx)}{\dg(dx)},
\\    \cH_{\nu}(\mu) - \mu^{x} H_{\nu}(x \mid \mu)
     &= - \mu^{x} \log \frac{\mu(dx)}{\dg(dx)} 
  \le -\mu(X) \log\frac{\mu(X)}{\dg(X)} \le 0.
 \end{align*}
This proves~\eqref{e.entropy-less-avg-algentr}.

Over a discrete space $\bX$, the function
$(x,\mu,\nu) \mapsto \frac{\mu(dx)}{\nu(dx)} = \frac{\mu(x)}{\nu(x)}$
is computable, therefore by the maximality property of
$H_{\nu}(x \mid \mu)$ we have 
$\frac{\mu(dx)}{\nu(dx)} \lem \t_{\nu}(x \mid \mu)$,
hence $\cH_{\nu}(\mu) = -\mu^{x} \log \frac{\mu(dx)}{\nu(dx)} 
 \gea \mu^{x} H_{\nu}(x \mid \mu)$.
 \end{proof} 

\subsection{Addition theorem}
The most important information-theoretical property of description
complexity is the following theorem (see for example~\cite{LiViBook97}):

 \begin{proposition}[Addition Theorem]\label{p.addition}
We have $H(x,y) \eqa H(x) + H(y \mid x, H(x))$.
 \end{proposition}

Mutual information is defined as $I(x : y) = H(x) + H(y) - H(x,y)$.
By the Addition theorem, we have
$I(x:y) \eqa H(y) - H(y \mid x,\, H(x)) \eqa H(x) - H(x \mid y,\,H(y))$.   
The two latter expressions show that in some sense, $I(x:y)$ is the
information held in $x$ about $y$ as well as the information held in $y$
about $x$.  
(The terms $H(x)$, $H(y)$ in the conditions are
logarithmic-sized corrections to this idea.)
Using~\eqref{e.ol-d}, it is interesting to view
mutual information $I(x : y)$ as a deficiency of randomness of
the pair $(x,y)$
in terms of the expression $\ol\d_{\mu}$, with respect to $\m \times \m$:
 \[
   I(x : y) = H(x) + H(y) - H(x,y) = \ol\d_{\m \times \m}(x, y).
 \]
Taking $\m$ as a kind of ``neutral'' probability, even if it is not quite
such, allows us to view $I(x:y)$ as a ``deficiency of independence''.
Is it also true that $I(x:y) \eqa \d_{\m \times \m}(x)$?
This would allow us to deduce, as Levin did, ``information conservation''
laws from randomness conservation laws.\footnote{We cannot use the 
test $\ol\t_{\mu}$ for this, since---as it can be shown easily--it 
does not obey randomness conservation.}

Expression $\d_{\m \times \m}(x)$ must be understood
again in the sense of compactification, as in
Section~\ref{s.neutral}.
There seem to be two reasonable ways to compactify the space
$\bbN\times\bbN$: we either compactify it directly, by adding a symbol
$\infty$, or we form the product $\ol\bbN \times \ol\bbN$.
With either of them, preserving
Theorem~\ref{t.test-charac-discr}, we would
have to check whether $H(x,y \mid \m \times \m) \eqa H(x,y)$.
But, knowing the function $\m(x)\times\m(y)$ we 
know the function $x \mapsto \m(x) \eqm \m(x) \times \m(0)$,
hence also the function $(x,y)\mapsto\m(x,y) = \m(\ang{x,y})$, where
$\ang{x,y}$ is any fixed computable pairing function.
Using this knowledge, it is possible to develop an argument similar to
the proof of Theorem~\ref{t.no-upper-semi-neutral}, showing that
$H(x,y \mid \m \times \m) \eqa H(x,y)$ does not hold.
 \begin{question} 
Is there a neutral measure $M$ with the property
$I(x:y) = \d_{M\times M}(x,y)$?
Is this true maybe for all neutral measures $M$?
If not, how far apart are the expressions $\d_{M\times M}(x,y)$ and
$I(x:y)$ from each other?
 \end{question}

The Addition Theorem (Proposition~\ref{p.addition})
can be generalized to the algorithmic entropy $H_{\mu}(x)$
introduced in~\eqref{e.alg-ent} (a somewhat similar generalization appeared
in~\cite{VovkVyugin93}).
The generalization, defining $H_{\mu,\nu} = H_{\mu\times\nu}$, is
 \begin{equation}\label{e.addition-general}
   H_{\mu,\nu}(x,y)\eqa
  H_\mu(x \mid \nu)+ H_\nu(y \mid x,\; H_\mu(x \mid \nu),\; \mu).
 \end{equation}
Before proving the general addition theorem, we establish a few useful
facts.

 \begin{proposition}\label{p.int.H.of.xy}
 We have
 \[
  H_{\mu}(x \mid \nu) \lea -\log \nu^{y} 2^{-H_{\mu,\nu}(x, y)}.
 \]
 \end{proposition}
 \begin{proof}
The function $f(x,\mu,\nu)$ that is the right-hand side, is upper
semicomputable by definition, and obeys $\mu^{x}2^{-f(x,\mu,\nu)} \le 1$.
Therefore the inequality follows from the minimum property of
$H_{\mu}(x)$.
 \end{proof}

Let us generalize the minimum property of $H_{\mu}(x)$.

 \begin{proposition}\label{p.univ-test-gener}
  Let $(x,y,\nu) \mapsto f_{\nu}(x,y)$ be a nonnegative lower semicomputable
function with $F_{\nu}(x) = \log \nu^{y} f_{\nu}(x,y)$.
Then for all $x$ with $F_{\nu}(x) > -\infty$ we have
 \[
   H_{\nu}(y \mid x, \flo{F_{\nu}(x)}) \lea -\log f_{\nu}(x,y) + F_{\nu}(x).
 \]
 \end{proposition}
 \begin{proof}
 Let us construct a lower semicomputable function 
$(x,y,m,\nu) \mapsto g_{\nu}(x,y,m)$ for
integers $m$ with the property that $\nu^{y} g_{\nu}(x,y,m) \le 2^{-m}$,
and for all $x$ with $F_{\nu}(x) \le -m$ we have
$g_{\nu}(x,y,m) = f_{\nu}(x,y)$.
Such a $g$ can be constructed by watching the approximation
of $f$ grow and cutting it off as soon as it would give $F_{\nu}(x) > -m$.
Now $(x,y,m,\nu) \mapsto 2^{m} g_{\nu}(x,y,m)$ is a uniform
conditional test of $y$ and hence it is $\lem 2^{-H_{\nu}(y \mid x, m)}$.
To finish the proof, substitute $-\flo{F_{\nu}(x)}$ for $m$ and rearrange.
 \end{proof}

Let $z \in \bbN$, then the inequality
 \begin{equation}\label{e.H.x.cond.z}
  H_{\mu}(x) \lea H(z) + H_{\mu}(x \mid z)
 \end{equation}
 will be a simple consequence of the general addition theorem.
The following lemma, needed in the proof of the theorem,
generalizes this inequality somewhat:

  \begin{lemma}\label{l.H.x.cond.z}
 For a com\-put\-able func\-tion $(y,z) \mapsto f(y,z)$ over $\bbN$,
we have
 \[
  H_{\mu}(x \mid y) \lea H(z) + H_{\mu}(x \mid f(y,z)).
 \]
 \end{lemma}
 \begin{proof}
  The function
 \[
  (x,y,\mu) \mapsto g_{\mu}(x, y)=\sum_{z} 2^{-H_{\mu}(x \mid f(y,z))-H(z)}
 \]
 is lower semicomputable,
and $\mu^{x} g_{\mu}(x, y) \le \sum_{z} 2^{-H(z)} \le 1$.
Hence $g_{\mu}(x, y) \lem 2^{-H_{\mu}(x \mid y)}$.
The left-hand side is a sum, hence the inequality holds for each
element of the sum: just what we had to prove.
 \end{proof}

As mentioned above, the
theory generalizes to measures that are not probability measures.
Taking $f_{\mu}(x,y)=1$ in Proposition~\ref{p.univ-test-gener} gives
the inequality
 \[
   H_{\mu}(x \mid \flo{\log \mu(X)}) \lea \log\mu(X),
 \]
with a physical meaning when $\mu$ is the phase space measure.
Using~\eqref{e.H.x.cond.z}, this implies
 \begin{equation}\label{e.unif.ub}
  H_\mu(x)\lea \log\mu(X) + H(\flo{\log\mu(X)}).
  \end{equation}

The following simple monotonicity property will be needed:

\begin{lemma}\label{l.mon}
 For $i < j$ we have
 \[
   i + H_\mu(x\mid i) \lea j + H_\mu(x\mid j) .
 \]
\end{lemma}
\begin{proof}
  From Lemma~\ref{l.H.x.cond.z}, with $f(i, n)=i + n$ we have
 \[
   H_{\mu}(x \mid i) - H_{\mu}(x \mid j) \lea H(j-i) \lea j-i.
 \]
\end{proof}

 \begin{theorem}[General addition]\label{t.addition-general}
The following inequality holds:
 \[
   H_{\mu,\nu}(x,y) \eqa
  H_{\mu}(x \mid \nu)+ H_{\nu}(y \mid x,\; H_\mu(x \mid \nu),\; \mu).
 \]
 \end{theorem}
 \begin{proof}
  To prove the inequality $\lea$, let us define
 \[
   G_{\mu,\nu}(x,y,m) =\min_{i\ge m}\;i +  H_{\nu}(y \mid x, i, \mu).
 \]
  Function $G_{\mu,\nu}(x,y,m)$ is upper semicomputable and decreasing
in $m$.
Therefore 
 \[
  G_{\mu,\nu}(x,y)  = G_{\mu,\nu}(x, y, H_{\mu}(x \mid \nu))
 \]
is also upper semicomputable since it is obtained by substituting an
upper semicomputable function for $m$ in $G_{\mu,\nu}(x,y,m)$.
Lemma~\ref{l.mon} implies 
 \begin{align*}
    G_{\mu,\nu}(x,y,m) &\eqa m +  H_{\nu}(y \mid x, m, \mu),
\\  G_{\mu,\nu}(x,y)   &\eqa H_{\mu}(x \mid \nu) + 
   H_{\nu}(y \mid x, H_{\mu}(x \mid \nu), \mu).
 \end{align*}
  Now, we have
 \begin{align*}
   \nu^{y} 2^{-m - H_{\nu}(y \mid x, m, \mu)} &\le 2^{-m},
\\   \nu^{y} 2^{-G_{\mu,\nu}(x,y)} &\lem 2^{-H_{\mu}(x \mid \mu)}.
 \end{align*}
 Therefore $\mu^{x}\nu^{y} 2^{-G} \lem 1$, implying
$H_{\mu,\nu}(x,y) \lea G_{\mu,\nu}(x,y)$ 
by the minimality property of $H_{\mu,\nu}(x,y)$.
This proves the $\lea$ half of our theorem.

To prove the inequality $\gea$, let
 \begin{align*}
    f_{\nu}(x,y,\mu) &= 2^{-H_{\mu,\nu}(x,y)},
\\ F_{\nu}(x, \mu)   &= \log \nu^{y} f_{\nu}(x,y,\mu).
 \end{align*}
  According to Proposition~\ref{p.univ-test-gener},
 \begin{align*}
  H_{\nu}(y \mid x,\flo{F},\mu) 
   &\lea -\log f_{\nu}(x,y,\mu) + F_{\nu}(x, \mu),
\\ H_{\mu,\nu}(x,y) &\gea -F + H_{\nu}(y \mid x, \cei{-F}, \mu).
 \end{align*}
 Proposition~\ref{p.int.H.of.xy} implies 
$-F_{\nu}(x, \mu) \gea H_{\mu}(x \mid \nu)$.
The monotonity lemma~\ref{l.mon} implies from here the $\gea$ half of the
theorem.
 \end{proof}

\subsection{Some special cases of the addition theorem; information}

  The function $H_{\mu}(\cdot)$ behaves quite differently for different
kinds of measures $\mu$.
Recall the following property of complexity:
 \begin{equation} \label{e.compl.of.fun}
  H(f(x)\mid y)\lea H(x\mid g(y)) \lea H(x) .
 \end{equation}
for any computable functions $f,g$
This implies
 \[
   H(y)\lea H(x,y).
 \]
  In contrast, if $\mu$ is a probability measure then
 \[
    H_{\nu}(y) \gea H_{\mu,\nu}(x, y).
 \]
  This comes from the fact that $2^{-H_{\nu}(y)}$ is a test for
$\mu\times\nu$.

Let us explore some of the consequences and meanings of the additivity
property.
As noted in~\eqref{e.Hmu-to-cond}, the
subscript $\mu$ can always be added to the condition: 
$H_{\mu}(x) \eqa H_{\mu}(x \mid \mu)$.
Similarly, we have
 \[
   H_{\mu,\nu}(x,y) := H_{\mu\times\nu}(x,y) 
  \eqa H_{\mu\times\nu}(x,y\mid \mu\times\nu)
  \eqa H_{\mu\times\nu}(x,y\mid \mu,\nu)
  =: H_{\mu,\nu}(x,y\mid \mu,\nu),
 \]
where only before-last inequality requires new (easy) consideration.

Let us assume that $X=Y=\Sg^{*}$, the discrete space of all strings.  
With
general $\mu,\nu$ such that $\mu(x),\nu(x) \ne 0$ for all $x$,
using~\eqref{e.alg-entr-charac}, the addition theorem specializes to the
ordinary addition theorem, conditioned on $\mu,\nu$:
 \[
   H(x,y\mid \mu,\nu) \eqa
  H(x \mid \mu,\nu)+ H(y \mid x,\; H(x \mid \mu,\nu),\; \mu,\nu).
 \]
In particular, whenever $\mu,\nu$ are computable, this is just the regular
addition theorem.

Just as above, we defined mutual information as 
$I(x : y) = H(x) + H(y) - H(x,y)$, the new addition theorem
suggests a more general definition
 \[
   I_{\mu,\nu}(x : y) = H_{\mu}(x \mid \nu) + 
   H_{\nu}(y\mid \mu) - H_{\mu,\nu}(x,y).
 \]
In the discrete case $X=Y=\Sg^{*}$ with everywhere positive
$\mu(x),\nu(x)$, this simplifies to
 \[
   I_{\mu,\nu}(x : y) = H(x \mid \mu,\nu) + H(y\mid \mu,\nu) 
                        - H(x,y | \mu,\nu),
 \]
which is $\eqa I(x:y)$ in case of computable $\mu,\nu$.
How different can it be for non-computable $\mu,\nu$?

In the general case, even for computable $\mu,\nu$, it seems
worth finding out how much this expression depends on the choice of
$\mu,\nu$.
Can one arrive at a general, natural definition of mutual information along
this path?

 \section{Conclusion}

When uniform randomness tests are defined in as general a
form as they were here, the theory of information conservation does not fit
nicely into the theory of randomness conservation as it did 
with~\cite{LevinUnif76} and~\cite{LevinRandCons84}.
Still, it is worth laying the theory onto broad
foundations that, we hope, can serve as a basis for further development.


\appendix

\section{Topological spaces}\label{s.top}

Given two sets $X,Y$, a \df{partial function} $f$ 
from $X$ to $Y$, defined on a subset of $Y$, will be denoted as
 \[
   f:\sbsq X \to Y.
 \]

\subsection{Topology}\label{ss.top}

A \df{topology} on a set $X$ is defined by a class $\tau$
of its subsets called \df{open sets}.
It is required that the empty set and $X$ are open, and that arbitrary
union and finite intersection of open sets is open.
The pair $(X, \tau)$ is called a \df{topological space}.
A topology $\tau'$ on $X$ is called \df{larger}, or \df{finer} than $\tau$
if $\tau' \spsq \tau$.
A set is called \df{closed} if its complement is open.
A set $B$ is called the \df{neighborhood} of a set $A$ if $B$ contains an
open set that contains $A$.
We denote by 
$\Cl{A}, \Int{A}$
the closure (the intersection of all closed sets containing $A$)
and the interior of $A$ (the union of all open sets in $A$) respectively.
Let
 \[
   \partial A = \Cl{A} \xcpt \Int{A}
 \]
denote the boundary of set $A$.
A \df{base} is a subset $\bg$ of $\tau$ such that every open set is the
union of some elements of $\bg$.
A \df{neighborhood} of a point is a base element containing it.
A \df{base of neighborhoods of a point} $x$ is a set $N$ of neighborhoods
of $x$ with the property that each neighborhood of $x$ contains an element
of $N$.
A \df{subbase} is a subset $\sg$ of $\tau$ such that every open set is the
union of finite intersections from $\sg$.

\begin{examples}\label{x.topol}\
 \begin{enumerate}[\upshape 1.]

  \item\label{i.x.topol.discr} 
Let $X$ be a set, and let $\bg$ be the set of all points of $X$.
The topology with base $\bg$ is the \df{discrete topology} of the
set $X$.
In this topology, every subset of $X$ is open (and closed).

  \item\label{i.x.topol.real}
Let $X$ be the real line $\bbR$, and let $\bg_{\bbR}$ be the set of
all open intervals $\opint{a}{b}$.
The topology $\tau_{\bbR}$ obtained from this base is the usual topology of
the real line.
When we refer to $\bbR$ as a topological space without qualification, 
this is the topology we will always have in mind.

  \item Let $X = \ol\bbR = \bbR \cup \{-\infty,\infty\}$, 
and let $\bg_{\ol\bbR}$ consist of all open intervals $\opint{a}{b}$
and in addition of all intervals of the forms $\lint{-\infty}{a}$ and
$\rint{a}{\infty}$.
It is clear how the space $\ol\bbR_{+}$ is defined.

  \item\label{i.x.topol.real-upper-converg}
Let $X$ be the real line $\bbR$.
Let $\bg_{\bbR}^{>}$ 
be the set of all open intervals $\opint{-\infty}{b}$.
The topology with base $\bg_{\bbR}^{>}$ is also a topology of the real
line, different from the usual one.
Similarly, let $\bg_{\bbR}^{<}$ 
be the set of all open intervals $\opint{b}{\infty}$.

  \item\label{i.x.topol.Cantor}
On the space $\Sg^{\og}$, let 
$\tg_{C} = \setof{A\Sg^{\og} : A \sbsq \Sg^{*}}$ be called the topology of
the \df{Cantor space} (over $\Sg$).

 \end{enumerate}
\end{examples}

A set is called a $G_{\dg}$ set if it is a countable intersection of open
sets, and it is an $F_{\sg}$ set if it is a countable union of closed sets.

For two topologies $\tau_{1},\tau_{2}$ over the same set $X$, we define
the topology $\tau_{1}\V \tau_{2} = \tau_{1} \cap \tau_{2}$, and
$\tau_{1} \et \tau_{2}$ as the smallest topology containing 
$\tau_{1} \cup \tau_{2}$.
In the example topologies of the real numbers above, we have
$\tau_{\bbR} = \tau_{\bbR}^{<} \et \tau_{\bbR}^{>}$.

We will always require the topology to have at least the $T_{0}$ property:
every point is determined by the class of open sets containing it.
This is the weakest one of a number of other possible separation
properties: both topologies of the real line in the example above have it.
A stronger such property would be the $T_{2}$ property:
a space is called a \df{Hausdorff} space, or $T_{2}$ space,
if for every pair of different
points $x,y$ there is a pair of disjoint open sets
$A,B$ with $x\in A$, $y\in B$.
The real line with topology $\tau_{\bbR}^{>}$ in
Example~\ref{x.topol}.\ref{i.x.topol.real-upper-converg} above is
not a Hausdorff space.
A space is Hausdorff if and only if every open set is the union of closed
neighborhoods.

Given two topological spaces $(X_{i}, \tau_{i})$ ($i=1,2$),
a function $f :\sbsq X_{1} \to X_{2}$ is
called \df{continuous} if for every open set $G \sbs X_{2}$ its inverse
image $f^{-1}(G)$ is also open.
If the topologies $\tau_{1},\tau_{2}$ are not clear from the context then
we will call the function $(\tau_{1}, \tau_{2})$-continuous.
Clearly, the property remains the same if we require it only for all
elements $G$ of a subbase of $X_{2}$.
If there are two continuous functions between $X$ and $Y$ that
are inverses of each other then the two spaces are called
\df{homeomorphic}.
We say that $f$ is continuous \df{at point} $x$ if for every neighborhood
$V$ of $f(x)$ there is a neighborhood $U$ of $x$ with $f(U) \sbsq V$.
Clearly, $f$ is continuous if and only if it is continuous in each point.

A \df{subspace} of a topological space $(X, \tau)$
is defined by a subset $Y \sbsq X$, and the topology
$\tau_{Y} = \setof{G \cap Y : G \in \tau}$, called the \df{induced}
topology on $Y$.
This is the smallest topology on $Y$
making the identity mapping $x \mapsto x$ continuous.
A partial function $f :\sbsq X \to Z$ with $\dom(f) = Y$
is continuous iff $f : Y \to Z$ is continuous.

For two topological spaces $(X_{i}, \tau_{i})$ ($i=1,2$), 
we define the \df{product topology}
on their product $X \times Y$: this is the
topology defined by the subbase
consisting of all sets $G_{1} \times X_{2}$ and all sets
$X_{1} \times G_{2}$ with $G_{i} \in \tau_{i}$.
The product topology is the smallest topology making the projection
functions $(x,y) \mapsto x$, $(x,y) \mapsto y$ continuous.
Given topological spaces $X,Y,Z$ we
call a two-argument function $f: X \times Y \mapsto Z$ continuous if
it is continuous as a function from $X \times Y$ to $Z$.
The product topology is defined similarly for
over the product $\prod_{i \in I} X_{i}$ of an arbitrary number of
spaces, indexed by some index set $I$.
We say that a function is
$(\tau_{1},\dots,\tau_{n},\mu)$-continuous if it is
$(\tau_{1} \times\dots\times\tau_{n},\mu)$-continuous.

 \begin{examples}\label{x.prod}\
  \begin{enumerate}[\upshape 1.]
  
   \item\label{i.x.prod.real}
The space $\bbR \times \bbR$ with the product
topology has the usual topology of the Euclidean plane.

   \item\label{i.x.top-inf-seq}
Let $X$ be a set with the discrete topology,
and $X^{\og}$ the set of infinite sequences with elements from $X$,
with the product topology.
A base of this topology is provided by all sets of the form $uX^{\og}$
where $u \in X^{*}$.
The elements of this base are closed as well as open.
When $X = \{0,1\}$ then this topology
is the usual topology of infinite binary sequences.

  \end{enumerate}
 \end{examples}

A real function $f : X_{1} \to \bbR$ is called continuous if it is
$(\tau_{1}, \tau_{\bbR})$-continuous.
It is called \df{lower semicontinuous} if it is 
$(\tau_{1}, \tau_{\bbR}^{<})$-continuous.
The definition of upper semicontinuity is similar.
Clearly, $f$ is continuous if and only if it is both lower and upper
semicontinuous.
The requirement of lower semicontinuity of $f$ is that
for each $r \in \bbR$, the set $\setof{x: f(x) > r}$ is open.
This can be seen to be equivalent to the requirement that the single set
$\setof{(x,r): f(x) > r}$ is open.
It is easy to see that the supremum of any set of lower semicontinuous
functions is lower semicontinuous.

Let $(X, \tau)$ be a topological space, and $B$ a subset of $X$.
An \df{open cover} of $B$ is a family of open sets whose union contains
$B$.
A subset $K$ of $X$ is said to be \df{compact} if every open cover of $K$
has a finite subcover.
Compact sets have many important properties: for example, a continuous
function over a compact set is bounded.

 \begin{example}\label{x.compact}\
 \begin{enumerate}[\upshape 1.]

  \item\label{i.compact.compactify}
 Every finite discrete space is compact.
An infinite discrete space $\bX = (X, \tau)$ is not compact, 
but it can be turned into a compact space $\ol\bX$ 
by adding a new element called $\infty$: let
$\ol X = X\cup\{\infty\}$, and 
$\ol\tau = \tau\cup\setof{\ol X \xcpt A: A \sbs X\txt{ closed }}$.
More generally, this simple operation can be performed with every space
that is \df{locally compact}, that each of its points has a compact
neighborhood.

  \item In a finite-dimensional Euclidean space, every bounded closed set
is compact.

  \item It is known that if $(\bX_{i})_{i\in I}$ is a family of compact
spaces then their direct product is also compact.

 \end{enumerate}
 \end{example}

A subset $K$ of $X$ is said to be \df{sequentially compact} 
if every sequence in $K$ has a convergent subsequence with limit in $K$.
The space is \df{locally compact} if every point has a compact
neighborhood.

\subsection{Metric spaces}\label{ss.metric}

In our examples for metric spaces, and later in our treatment of the space
of probability measures, we refer to~\cite{BillingsleyConverg68}.
A \df{metric space} is given by a set $X$ and a distance function 
$d : X\times X \to \bbR_{+}$ satisfying the
\df{triangle inequality} $d(x, z) \le d(x, y) + d(y, z)$ and also
property that $d(x,y) = 0$ implies $x = y$.
For $r \in\bbR_{+}$, the sets 
 \[
  B(x,r) = \setof{y : d(x, y) < r},\quad
  \ol B(x,r) = \setof{y : d(x, y) \le r}
 \]
are called the open and closed \df{balls} with radius $r$ and center $x$.
A metric space is also a topological space, with the base that is the set
of all open balls.
Over this space, the distance function $d(x,y)$ is obviously continuous.
Each metric space is a Hausdorff space; moreover, it has the
following stronger property.
For every pair of different points $x,y$ there is a continuous function 
$f : X \to \bbR$  with $f(x) \ne f(y)$.
(To see this, take $f(z) =  d(x, z)$.)
This is called the $T_{3}$ property.
A metric space is \df{bounded} when $d(x,y)$ has an upper bound on $X$.
A topological space is called \df{metrizable} if its topology can be
derived from some metric space.

\begin{notation}
For an arbitrary set $A$ and point $x$ let 
\begin{align}\nonumber
    d(x, A) &= \inf_{y \in A} d(x,y),
\\\label{e.Aeps}
  A^{\eps} &= \setof{x : d(x, A) < \eps}.
  \end{align}
\end{notation}

 \begin{examples}\label{x.metric}\
 \begin{enumerate}[\upshape 1.]

  \item The real line with the distance $d(x,y) = |x - y|$ is a metric
space.
The topology of this space is the usual topology $\tau_{\bbR}$ of the real
line.

  \item The space $\bbR \times \bbR$ with the Euclidean distance gives the
same topology as the product topology of $\bbR \times \bbR$.

  \item An arbitrary set $X$ with the distance $d(x,y)=1$ for all pairs
$x,y$ of different elements, is a metric space that induces the discrete
topology on $X$.

  \item\label{i.x.metric-inf-seq} 
Let $X$ be a bounded metric space, and let
$Y = X^{\og}$ be the set of infinite sequences 
$x = (x_{1}, x_{2}, \dotsc)$
with distance function $d^{\og}(x,y) = \sum_{i} 2^{-i} d(x_{i},y_{i})$.
The topology of this space is the same as the product topology defined 
in Example~\ref{x.prod}.\ref{i.x.top-inf-seq}.

  \item\label{i.x.metric-nonsep} 
Let $X$ be a metric space, and let
$Y = X^{\og}$ be the set of infinite bounded sequences 
$x = (x_{1}, x_{2}, \dotsc)$
with distance function $d(x, y) = \sup_{i} d(x_{i}, y_{i})$.

  \item\label{i.x.C(X)}
Let $X$ be a metric space, and let
$C(X)$ be the set of bounded continuous functions over $X$ with
distance function $d'(f, g) = \sup_{x} d(f(x), g(x))$.
A special case is $C\clint{0}{1}$ where the interval $\clint{0}{1}$ of real
numbers has the usual metric.

  \item\label{i.x.l2}
Let $l_{2}$ be the set of infinite sequences $x = (x_{1}, x_{2}, \dotsc)$
of real numbers with the property that $\sum_{i} x_{i}^{2} < \infty$.
The metric is given by 
the distance $d(x, y) = (\sum_{i} |x_{i} - y_{i}|^{2})^{1/2}$.

 \end{enumerate}
 \end{examples}

A topological space has the \df{first countability property} if each point
has a countable base of neighborhoods.
Every metric space has the first countability property since we can
restrict ourselves to balls with rational radius.
Given a topological space $(X, \tau)$ and a sequence $x = (x_{1}, x_{2},
\dotsc)$ of elements of $X$, we say that $x$ \df{converges} to a point $y$
if for every neighborhood $G$ of $y$ there is a $k$ such that for all 
$m > k$ we have $x_{m} \in G$.
We will write $y = \lim_{n \to \infty} x_{n}$.
It is easy to show that if spaces $(X_{i}, \tau_{i})$ $(i=1,2)$
have the first countability property then a function $f : X \to Y$ is
continuous if and only if for every convergent sequence $(x_{n})$ we have
$f(\lim_{n} x_{n}) = \lim_{n} f(x_{n})$.
A topological 
space has the \df{second countability property} if the whole space
has a countable base.
For example, the space $\bbR$ has the second countability property
for all three topologies $\tau_{\bbR}$, $\tau_{\bbR}^{<}$, $\tau_{\bbR}^{>}$.
Indeed, we also
get a base if instead of taking all intervals, we only take
intervals with rational endpoints.
On the other hand, the metric space of 
Example~\ref{x.metric}.\ref{i.x.metric-nonsep} does not have
the second countability property.
In a topological space $(X, \tau)$, a set $B$ of points is called
\df{dense} at a point $x$ if it intersects every neighborhood of $x$.
It is called \df{everywhere dense}, or \df{dense}, if it is dense at every
point.
A metric space is called \df{separable} if it has a countable everywhere
dense subset.
This property holds if and only if the space as a topological
space has the second countability property.

  \begin{example}\label{x.Cclint{0}{1}}
In Example~\ref{x.metric}.\ref{i.x.C(X)}, for $X=\clint{0}{1}$, we can
choose as our everywhere dense set the set of all polynomials with rational
coefficients, or alternatively,
the set of all piecewise linear functions whose graph has
finitely many nodes at rational points.
  \end{example}

Let $X$ be a metric space, and let
$C(X)$ be the set of bounded continuous functions over $X$ with
distance function $d'(f, g) = \sup_{x} d(f(x), g(x))$.
A special case is $C\clint{0}{1}$ where the interval $\clint{0}{1}$ of real
numbers has the usual metric.

Let $(X, d)$ be a metric space, and $a = (a_{1}, a_{1},\dotsc)$ an infinite
sequence.
A metric space is called \df{complete} if every Cauchy sequence in it has a
limit.
It is well-known that every metric space can be embedded (as an everywhere
dense subspace) into a complete space.

It is easy to see that in a metric space, every closed set is a
$G_{\dg}$ set (and every open set is an $F_{\sg}$ set).

 \begin{example}
Consider the set $D\clint{0}{1}$ of functions over
$\clint{0}{1}$ that are right continuous and have left limits everywhere.
The book~\cite{BillingsleyConverg68} introduces two different metrics for
them: the Skorohod metric $d$ and the $d_{0}$ metric.
In both metrics, two functions are close if a slight monotonic continuous
deformation of the coordinate makes them uniformly close.
But in the $d_{0}$ metric, the slope of the deformation must be close to 1.
It is shown that the two metrics give rise to the same topology;
however, the space with metric $d$ is not complete, and the 
space with metric $d_{0}$ is.
 \end{example}

Let $(X, d)$ be a metric space. 
It can be shown that a subset $K$ of $X$ is compact if and only 
if it is sequentially compact.
Also, $K$ is compact if and only if it is closed and
for every $\eps$, there is a finite set of
$\eps$-balls (balls of radius $\eps$) covering it.

We will develop the theory of randomness over separable
complete metric spaces.
This is a wide class of spaces encompassing most spaces of practical
interest.
The theory would be simpler if we restricted it to compact or locally
compact spaces; however, some important spaces like $C\clint{0}{1}$
(the set of continuouos functions over the interval $\clint{0}{1}$, with
the maximum difference as their distance) are not locally compact.

Given a function $f: X \to Y$ between metric spaces and $\bg > 0$,
let $\Lip_{\bg}(X,Y)$ denote the
set of functions (called the Lip\-schitz$(\bg)$ functions, or simply
Lip\-schitz functions) satisfying
 \begin{equation}\label{e.Lipschitz}
   d_{Y}(f(x), f(y)) \le \bg d_{X}(x, y).
 \end{equation}
All these functions are uniformly continuous.
Let $\Lip(X) = \Lip(X,\bbR) = \bigcup_{\bg} \Lip_{\bg}$ 
be the set of real Lip\-schitz functions over $X$.

\section{Measures}\label{s.measures}

For a survey of measure theory, see for example~\cite{PollardUsers01}.

\subsection{Set algebras}

A (Boolean set-) \df{algebra} is a set of subsets of some set $X$
closed under intersection and complement (and then, of course, under
union).
It is a \df{$\sg$-algebra} if it is also closed 
under countable intersection 
(and then, of course, under countable union).
A \df{semialgebra} is a set $\cL$
of subsets of some set $X$ closed under
intersection, with the property that the complement of every element 
of $\cL$ is the disjoint union of a finite number of elements of $\cL$.
If $\cL$ is a semialgebra then the set of finite unions of elements of
$\cL$ is an algebra.

 \begin{examples}\label{x.algebras}\
  \begin{enumerate}[\upshape 1.]

   \item\label{i.x.algebras.l-closed}
The set $\cL_{1}$ of left-closed intervals of the line (including intervals
of the form $\opint{-\infty}{a}$) is a semialgebra.

   \item
The set $\cL_{2}$ of all intervals of the line
(which can be open, closed, left-closed or
right-closed), is a semialgebra.

  \item\label{i.x.algebras.inf-seq}
In  the set $\{0,1\}^{\og}$ of infinite 0-1-sequences, the
set $\cL_{3}$ of all subsets of the form $u\{0,1\}^{\og}$ with
$u\in\{0,1\}^{*}$, is a semialgebra.

   \item
The $\sg$-algebra $\cB$ generated by $\cL_{1}$, is the same as the one
generated by $\cL_{2}$, and is also the same as the one generated by the
set of all open sets: it is called the family of \df{Borel sets} of the
line.
The Borel sets of the extended real line $\ol\bbR$ are defined similarly.
  
   \item
Given $\sg$-algebras $\cA,\cB$ in sets $X,Y$, the product $\sg$-algebra
$\cA\times\cB$ in the space $X \times Y$ is the one generated by all
elements $A \times Y$ and $X \times B$ for $A\in\cA$ and $B\in\cB$.

  \end{enumerate}
 \end{examples}

\subsection{Measures}\label{ss.measures}

A \df{measurable space} is a pair $(X, \cS)$ where $\cS$ is a $\sg$-algebra
of sets of $X$.
A \df{measure} on a measurable space $(X, \cS)$ is a function 
$\mu : B \to \ol\bbR_{+}$ that is \df{$\sg$-additive}:
this means that for every countable family $A_{1}, A_{2},\dotsc$ of
disjoint elements of $\cS$ we have 
$\mu(\bigcup_{i} A_{i}) = \sum_{i} \mu(A_{i})$.
A measure $\mu$ is \df{$\sg$-finite} if the whole space is the union of a
countable set of subsets whose measure is finite.
It is \df{finite} if $\mu(X) < \infty$. 
It is a \df{probability measure} if $\mu(X) = 1$.

It is important to understand how a measure can be defined in practice.
Algebras are generally simpler to grasp constructively
than $\sg$-algebras; semialgebras are yet simpler.
Suppose that $\mu$ is defined over a semialgebra $\cL$ and is additive.
Then it can always be uniquely extended to an additive function over
the algebra generated by $\cL$.
The following is an important theorem of measure theory.

 \begin{proposition}\label{p.Caratheo-extension}
Suppose that a nonnegative set function defined over a semialgebra $\cL$
is $\sg$-additive.
Then it can be extended uniquely to the $\sg$-algebra generated by $\cL$.
 \end{proposition}

 \begin{examples}\label{x.semialgebra}\
  \begin{enumerate}[\upshape 1.]

   \item Let $x$ be point and let $\mu(A) = 1$ if $x \in A$ and $0$
otherwise.
In this case, we say that $\mu$ is \df{concentrated} on the point $x$.

   \item\label{i.left-cont} Consider the the line $\bbR$, and the
algebra $\cL_{1}$ defined 
in Example~\ref{x.algebras}.\ref{i.x.algebras.l-closed}.
Let $f : \bbR \to \bbR$ be a monotonic real function.
We define a set function over $\cL_{1}$ as follows.
Let $\lint{a_{i}}{b_{i}}$, ($i=1,\dots,n$) be a set of disjoint left-closed
intervals.
Then $\mu(\bigcup_{i} \lint{a_{i}}{b_{i}}) = \sum_{i} f(b_{i}) - f(a_{i})$. 
It is easy to see that $\mu$ is additive.
It is $\sg$-additive if and only if $f$ is left-continuous.

  \item\label{i.measure-Cantor} Let $B = \{0,1\}$, and consider the set 
$B^{\og}$ of infinite 0-1-sequences, and the
semialgebra $\cL_{3}$ of
Example~\ref{x.algebras}.\ref{i.x.algebras.inf-seq}.
Let $\mu : B^{*} \to \bbR^{+}$ be a function.
Let us write $\mu(uB^{\og}) = \mu(u)$ for all $u \in B^{*}$.
Then it can be shown that the following conditions are equivalent:
$\mu$ is $\sg$-additive over $\cL_{3}$; it is 
additive over $\cL_{3}$; the equation $\mu(u) = \mu(u0) + \mu(u1)$ 
holds for all $u \in B^{*}$.

  \item The nonnegative linear combination of any finite number of measures
is also a measure.
In this way, it is easy to construct arbitrary measures concentrated on a
finite number of points.

  \item Given two measure spaces $(X,\cA,\mu)$ and $(Y,\cB,\nu)$ it is
possible to 
define the product measure $\mu\times\nu$ over the measureable space
$(X\times Y, \cA\times\cB)$.
The definition is required to satisfy
 $\mu\times\nu(A\times B) = \mu(A)\times\nu(B)$, and is determined uniquely
by this condition.
If $\nu$ is a probability measure then, of course,
 $\mu(A) = \mu\times\nu(A \times Y)$.

  \end{enumerate}
 \end{examples}

 \begin{remark}\label{r.measure.step-by-step}
Example~\ref{x.semialgebra}.\ref{i.measure-Cantor} shows a particularly
attractive way to define measures.
Keep splitting the values $\mu(u)$ in an arbitrary way into
$\mu(u0)$ and $\mu(u1)$, and the resulting values on the semialgebra define
a measure.
Example~\ref{x.semialgebra}.\ref{i.left-cont} is less attractive,
since in the process of defining $\mu$ on all intervals and only keeping
track of finite additivity, we may end up with
a monotonic function that is not left continuous, and thus with a measure
that is not $\sg$-additive.
In the subsection on probability measures over a metric space, we will find
that even on the real line, there is a way to define measures in a
step-by-step manner, and only checking for consistency along the way.
 \end{remark} 

A \df{probability space} is a triple $(X, \cS, P)$ where $(X, \cS)$ is a
measurable space and $P$ is a probability measure over it.

Let $(X_{i}, \cS_{i})$ ($i=1,2$) be measurable spaces, and let 
$f : X \to Y$ be a mapping.
Then $f$ is \df{measurable} if and only if for each element $B$ of
$\cS_{2}$, its inverse image $f^{-1}(B)$ is in $\cS_{1}$.
If $\mu_{1}$ is a measure over $(X_{1}, \cS_{1})$ then
$\mu_{2}$ defined by $\mu_{2}(A) = \mu_{1}(f^{-1}(A))$ is a measure over
$X_{2}$ called the measure \df{induced} by  $f$.

\subsection{Integral}\label{ss.integral}

A measurable function $f : X \to \bbR$ is called a \df{step function} if
its range is finite.
The set of step functions is closed with respect to linear combinations and
also with respect to the operations $\et,\V$.
Such a set of functions is called a \df{Riesz space}.


Given a step function which takes values $x_{i}$ on sets $A_{i}$, and a
finite measure $\mu$, we define 
 \[
 \mu(f) = \mu f = \int f\,d\mu = \int f(x) \mu(d x) 
   = \sum_{i} x_{i} \mu(A_{i}).
 \]
This is a linear positive functional on the set of step functions.
Moreover, it can be shown that it
is continuous on monotonic sequences: if $f_{i} \searrow 0$
then $\mu f_{i} \searrow 0$.
The converse can also be shown:
Let $\mu$ be a linear positive functional on step functions 
that is continuous on monotonic sequences.
Then the set function $\mu(A) = \mu(1_{A})$ is a finite measure.

 \begin{proposition}\label{p.Riesz-extension}
Let $\cE$ be any Riesz space of functions with the property that
$1 \in \cE$.
Let $\mu$ be a positive linear functional on $\cE$ continuous on monotonic
sequences, with $\mu 1 = 1$.
The functional $\mu$ can be extended to the set 
$\cE_{+}$ of monotonic limits of nonnegative elements of $\cE$, by
continuity.
In case when $\cE$ is the set of all step functions, the set $\cE_{+}$ is
the set of all nonnegative measurable functions.
 \end{proposition}

Let us fix a finite measure $\mu$ over a measurable space $(X, \cS)$.
A measurable function $f$ is called \df{integrable} with respect to $\mu$
if $\mu |f|^{+} < \infty$ and $\mu |f|^{-} < \infty$.
In this case, we define $\mu f = \mu |f|^{+} - \mu |f|^{-}$.
The set of integrable functions is a Riesz space, and the positive linear
functional $\mu$ on it is continuous with respect to monotonic sequences.
The continuity over monotonic sequences also implies the following
\df{bounded convergence theorem}.

 \begin{proposition}
Suppose that functions $f_{n}$ are integrable and
$|f_{n}| < g$ for some integrable function $g$.
Then $f = \lim_{n} f_{n}$ is integrable and $\mu f = \lim_{n} \mu f_{n}$.
 \end{proposition}

Two measurables functions $f,g$ are called \df{equivalent} with respect to
$\mu$ if $\mu(f - g) = 0$.

For two-dimensional integration, the following theorem holds.

 \begin{proposition}
Suppose that function $f(\cdot,\cdot)$ is integrable over
the space $(X\times Y, \cA\times\cB, \mu\times\nu)$.
Then for $\mu$-almost all $x$, the function $f(x,\cdot)$ is integrable over
$(Y,\cB,\nu)$, and the function $x\mapsto\nu^{y}f(x,y)$ 
is integrable over $(X,\cA,\mu)$
with $(\mu\times\nu) f = \mu^{x}\mu^{y}f$.
 \end{proposition}

\subsection{Density}

Let $\mu, \nu$ be two measures over the same measurable space.
We say that $\nu$ is \df{absolutely continuous} with respect to $\mu$, or
that $\mu$ \df{dominates} $\nu$, if
for each set $A$, $\mu(A) = 0$ implies $\nu(A) = 0$.
It can be proved that this condition is equivalent to the condition that
there is a positive real number $c$ with $\nu \le c \mu$.
Every nonnegative integrable function $f$ defines a new measure $\nu$ via
the formula $\nu(A) = \mu(f\cdot 1_{A})$.
This measure $\nu$ is absolutely continuous with respect to $\mu$.
The Radon-Nikodym theorem says that the converse is also true.

 \begin{proposition}[Radon-Nikodym theorem]
If $\nu$ is dominated by $\mu$ then there is a nonnegative integrable
function $f$ such that $\nu(A) = \mu(f \cdot 1_{A})$ for all measurable
sets $A$.
The function $f$ is defined uniquely to within equivalence with respect to
$\mu$.
 \end{proposition}

The function $f$ of the Radom-Nikodym Theorem above
is called the \df{density} of $\nu$ with respect to $\mu$.
We will denote it by
 \[
   f(x) = \frac{\mu(dx)}{\nu(dx)} = \frac{d\mu}{d\nu}.
 \]
The following theorem is also standard.

 \begin{proposition}\label{p.density-props}\
 \begin{enumerate}[(a)]
  \item
Let $\mu, \nu, \eta$ be measures such that $\eta$ is absolutely continuous
with respect to $\mu$ and $\mu$ is absolutely continuous with respect to
$\nu$.
Then the ``chain rule'' holds:
 \begin{equation}\label{e.chain-rule}
  \frac{d\eta}{d\nu} = \frac{d\eta}{d\mu} \frac{d\mu}{d\nu}.
 \end{equation}

  \item
If $\frac{\nu(dx)}{\mu(dx)} > 0$ for all $x$ then
$\mu$ is also absolutely continuous with respect to $\nu$ and
$\frac{\mu(dx)}{\nu(dx)} = \Paren{\frac{\nu(dx)}{\mu(dx)}}^{-1}$.
 \end{enumerate}
 \end{proposition}

Let $\mu, \nu$ be two measures, then both are dominated by some measure
$\eta$ (for example by $\eta = \mu + \nu$).
Let their densities with respect to $\eta$ be $f$ and $g$.
Then we define the \df{total variation distance} of the two measures
as
 \[
  D(\mu, \nu)=\eta(|f - g|).
 \]
It is independent of the dominating measure $\eta$.

 \begin{example}
Suppose that the space $X$ can be partitioned into 
disjoint sets $A,B$ such that $\nu(A)=\mu(B) = 0$.
Then $D(\mu, \nu) = \mu(A) + \nu(B) = \mu(X) + \nu(X)$.
 \end{example}

\subsection{Random transitions}\label{ss.transitions}

Let $(X, \cA)$, $(Y, \cB)$ be two measureable spaces (defined in
Subsection~\ref{ss.measures}).
We follow the definition given in~\cite{PollardUsers01}.
Suppose that a family of probability
measures $\Lg = \setof{\lg_{x} : x \in X}$ on $\cB$ is given.
We call it a \df{probability kernel}, (or Markov kernel, or conditional
distribution) if the map $x \mapsto \lg_{x} B$ is measurable for each
$B \in \cB$.
When $X,Y$ are finite sets then $\lg$ is a Markov transition matrix.
The following theorem shows that $\lg$ assigns a joint distribution over
the space $(X \times Y, \cA\times\cB)$ to each input distribution $\mu$.

 \begin{proposition} For each nonnegative $\cA\times \cB$-measureable
function $f$ over $X \times Y$,
 \begin{enumerate}[\upshape 1.]
  \item the function $y \to f(x,y)$ is $\cB$-measurable for each fixed $x$;
  \item $x \to \lg_{x}^{y} f(x, y)$ is $\cA$-measurable;
  \item the integral $f \to \mu^{x} \lg_{x}^{y} f(x, y)$ defines
a measure on $\cA \times \cB$.
 \end{enumerate}
 \end{proposition}

According to this proposition, given a probability kernel $\Lg$,
to each measure $\mu$ over $\cA$ corresponds a measure over
$\cA \times \cB$.
We will denote its marginal over $\cB$ as
 \begin{equation}\label{e.Markov-op-meas}
   \Lg^{*} \mu.
 \end{equation}
For every measurable function $g(y)$ over $Y$, we can define the measurable 
function $f(x) = \lg_{x} g = \lg_{x}^{y} g(y)$: we write
 \begin{equation}\label{e.Markov-op-fun}
   f = \Lg g.
 \end{equation}
The operator $\Lg$ is linear, and monotone with $\Lg 1 = 1$.
By these definitions, we have
 \begin{equation}\label{e.Lg-Lg*}
  \mu(\Lg g) = (\Lg^{*}\mu) g.
 \end{equation}

 \begin{example}\label{x.determ-trans}
Let $h : X \to Y$ be a measureable function, and
let $\lg_{x}$ be the measure $\dg_{h(x)}$ concentrated on the
point $h(x)$.
This operator, denoted $\Lg_{h}$ is, in fact, a deterministic transition,
and we have $\Lg_{h} g = g \circ h$.
In this case, we will simplify the notation as follows:
 \[
   h^{*}\mu =  \Lg_{h}^{*}.
 \]
 \end{example}

\subsection{Probability measures over a metric
space}\label{ss.measure-metric}

We follow the exposition of~\cite{BillingsleyConverg68}.
Whenever we deal with probability measures on a metric space, we will
assume that our metric space is complete and separable (Polish).
Let $\bX = (X, d)$ be a complete separable metric space.
It gives rise to a measurable space, where the measurable sets are the
Borel sets of $\bX$.
It can be shown that, if $A$ is a Borel set and $\mu$ is a finite measure
then there are sets
$F \sbsq A \sbsq G$ where $F$ is an $F_{\sg}$ set, $G$ is a $G_{\dg}$ set,
and $\mu(F) = \mu(G)$.
Let $\cB$ be a base of open sets closed under intersections.
Then it can be shown that $\mu$ is determined by its values on elements of
$\cB$.
The following proposition follows then essentially from
Proposition~\ref{p.Caratheo-extension}.

 \begin{proposition}\label{p.Caratheo-topol}
Let $\cB^{*}$ be the set algebra generated by the above base
$\cB$, and let $\mu$ be any
$\sigma$-additive set function on $\cB^{*}$ with $\mu(X)=1$.
Then $\mu$ can be extended uniquely to a probability measure.
 \end{proposition}

We say that a set $A$ is a \df{continuity set} of measure $\mu$ if
$\mu(\partial A) = 0$: the boundary of $A$ has measure 0.

\subsubsection{Weak topology}
Let 
 \[
   \cM(\bX)
 \]
be the set of probability measures on the metric space $\bX$.
Let 
 \[
   \dg_{x}
 \]
be a probability measure concentrated on point $x$.
Let $x_{n}$ be a sequence of points converging to point $x$ but with 
$x_{n} \ne x$.
We would like to say that $\dg_{x_{n}}$ converges to $\dg_{x}$.
But the total variation distance $D(\dg_{x_{n}}, \dg_{x})$ is 2
for all $n$.
This suggests that the total variation distance is not generally the best
way to compare probability measures over a metric space.
We say that a sequence of probability
measures $\mu_{n}$ over a metric space $(X, d)$
\df{weakly converges} to measure $\mu$ if for all bounded continuous
real functions $f$ over $X$ we have $\mu_{n} f \to \mu f$.
This \df{topology of weak convergence} $(\cM, \tau_{w})$
can be defined using a number of different subbases.
The one used in the original definition
is the subbase consisting of all sets of the form
 \[
   A_{f,c} = \setof{\mu : \mu f < c}
 \]
for bounded continuous functions $f$ and real numbers $c$.
We also get a subbase (see for example~\cite{PollardUsers01}) 
if we restrict ourselves to the set
$\Lip(X)$ of Lip\-schitz functions defined in~\eqref{e.Lipschitz}.
Another possible subbase giving rise to the same topology
consists of all sets of the form
 \begin{equation}\label{e.measure-on-open}
   B_{G,c} = \setof{\mu : \mu(G) > c}
 \end{equation}
for open sets $G$ and real numbers $c$.
Let us find some countable subbases.
Since the space $\bX$ is separable, there is a sequence $U_{1}, U_{2},
\dotsc$ of open sets that forms a base.
We can restrict the subbase of the space of measures to those sets 
$B_{G, c}$ where $G$ is the union of a finite number of base elements
$U_{i}$ and $c$ is rational.
Thus, the space $(\cM, \tau_{w})$ itself has the second countability
property.

It is more convenient to define a countable subbase using bounded
continuous functions $f$, since 
$\mu \mapsto \mu f$ is continuous on such functions, while 
$\mu \mapsto \mu U$ is typically not continuous when $U$ is an open set.
Let $\cF_{0}$ be the set of functions introduced before~\eqref{e.bd-Lip-seq}.
Let 
  \[
  \cF_{1}
  \]
be the set of functions $f$ with the property that $f$ is the
minimum of a finite number of elements of $\cF_{0}$.
Note that each element $f$ of $\cF_{1}$ is bounded between 0 and 1, and
from
its definition, we can compute a bound $\bg$ such that $f\in\Lip_{\bg}$.

 \begin{proposition}\label{p.Portmanteau}
The following conditions are equivalent:
 \begin{enumerate}[\upshape 1.]
  \item $\mu_{n}$ weakly converges to $\mu$.
  \item $\mu_{n} f \to \mu f$ for all $f \in \cF_{1}$.
  \item For every Borel set $A$, that is a continuity set of $\mu$, we have
$\mu_{n}(A) \to \mu(A)$.
  \item For every closed set $F$, $\lim\inf_{n} \mu_{n}(F) \ge \mu(F)$.
  \item For every open set $G$, $\lim\sup_{n} \mu_{n}(G) \le \mu(G)$.
 \end{enumerate}
 \end{proposition}

As a subbase
 \begin{equation}\label{e.metric-measure-subbase}
   \sg_{\cM}
 \end{equation}
for $\cM(x)$, we choose the sets
$\setof{\mu : \mu f < r}$ and $\setof{\mu : \mu f > r}$ for all 
$f \in \cF_{1}$ and $r \in \bbQ$.
Let $\cE$
be the set of functions introduced in~\eqref{e.bd-Lip-seq}.
It is a Riesz space as defined in Subsection~\ref{ss.integral}.
A reasoning combining Propositions~\ref{p.Caratheo-extension} 
and~\ref{p.Riesz-extension} gives the following.

 \begin{proposition}\label{p.metric-Riesz-extension}
Suppose that a positive linear functional $\mu$ with $\mu 1 = 1$ is defined
on $\cE$ 
that is continuous with respect to monotone convergence.
Then $\mu$ can be extended uniquely to a probability
measure over $\bX$ with $\mu f = \int f(x) \mu(dx)$ for all $f \in \cE$.
 \end{proposition}

\subsubsection{Prokhorov distance}\label{sss.Prokh}
The definition of measures in the style of
Proposition~\ref{p.metric-Riesz-extension}
is not sufficiently constructive.
Consider a gradual definition of the measure $\mu$, extending it
to more and more elements of $\cE$, while keeping the positivity and
linearity property.
It can happen that the function $\mu$ we end up with in the limit, is not
continuous with respect to monotone convergence.
Let us therefore metrize the space of
measures: then an arbitrary measure can be defined as the limit
of a Cauchy sequence of simple meaures.

One metric that generates the topology of weak convergence is
the \df{Prokhorov distance} $p(\mu, \nu)$:
the infimum of all those $\eps$ for which, for all Borel sets $A$ we
have (using the notation~\eqref{e.Aeps})
 \[
   \mu(A) \le \nu(A^{\eps}) + \eps.
 \]
It can be shown that this is a distance and it generates the weak
topology.
The following result helps visualize this distance:

 \begin{proposition}[Coupling Theorem, see~\protect\cite{Strassen65}]
\label{p.coupling}
Let $\mu,\nu$ be two probability measures over a complete separable metric
space $\bX$ with $p(\mu, \nu) \le\eps$.
Then there is a probability measure $P$ on the space $\bX \times \bX$
with marginals $\mu$ and $\nu$ such that for a pair of random variables
$(\xi,\eta)$
having joint distribution $P$ we have
 \[
   P\setof{d(\xi,\eta) > \eps} \le \eps.
 \]
 \end{proposition}

Since this topology has the second countability property, 
the metric space defined by the distance $p(\cdot,\cdot)$ is separable.
This can also be seen directly.
Let $S$ be a countable everywhere dense set of points in $X$.
Consider the set of $\cM_{0}(X)$ of
those probability measures that are concentrated on finitely many points of
$S$ and assign rational values to them.
It can be shown that $\cM_{0}(X)$ is everywhere dense in the metric space
$(\cM(X), p)$; so, this space is separable.
It can also be shown that $(\cM(X), p)$ is complete.
Thus, a measure can be given as the limit of a sequence of elements
$\mu_{1},\mu_{2},\dots$ of
$\cM_{0}(X)$, where $p(\mu_{i},\mu_{i+1}) < 2^{-i}$.

The definition of the Prokhorov distance quantifies over all Borel sets.
However, in an important simple case, it can be handled efficiently.

 \begin{proposition}\label{p.simple-Prokhorov-ball}
Assume that measure $\nu$ is concentrated on 
a finite set of points $S \sbs X$.
Then the condition $p(\nu,\mu) < \eps$ is equivalent to
the finite set of conditions
 \begin{equation}\label{e.finite-Prokhorov}
   \mu(A^{\eps}) > \nu(A) - \eps
 \end{equation}
for all $A \sbs S$.
 \end{proposition}

\subsubsection{Relative compactness}
A set $\Pi$
of measures in $(\cM(X), p)$ is called \df{relatively compact} if every
sequence of elements of $\Pi$ contains a convergent subsequence.
Relative compactness is an important property for proving convergence of
measures.
It has a useful characterization.
A set of $\Pi$ of measures is called \df{tight} if for every $\eps$ there
is a compact set $K$ such that $\mu(K) > 1 - \eps$ for all $\mu$ in $\Pi$.
Prokhorov's theorem  states (under our assumptions of the separability and
completeness of $(X, d)$) that a set of measures is relatively compact if
and only if it is tight and if and only if its closure is compact
in $(\cM(X), p)$.
In particular, the following fact is known.

  \begin{proposition}\label{p.measures-compact}
The space $(\cM(\bX), p)$ of measures is compact if and
only if the space $(X, d)$ is compact.
  \end{proposition}

So, if $(X, d)$ is not compact then the set of measures is not compact.
But still, each measure $\mu$
is ``almost'' concentrated on a compact set.
Indeed, the one-element set $\{\mu\}$ is compact and therefore
by Prokhorov's theorem tight.
Tightness says that for each $\eps$ a mass of size $1-\eps$ of $\mu$ is
concentrated on some compact set.

\section{Computable analysis}

If for some finite or infinite sequences $x,y,z,w$, we have
$z = wxy$ then we write $w \sqsbsq  z$ ($w$ is a \df{prefix} of $z$) and 
$x \ltri z$.
For integers, we will use the toupling functions
 \[
   \ang{i, j} = \frac{1}{2} (i+1)(i+j+1) + j,
\quad \ang{n_{1},\dots,n_{k+1}} = \ang{\ang{n_{1},\dots,n_{k}},n_{k+1}}.
 \]
Inverses: $\pi_{i}^{k}(n)$.

Unless said otherwise, the alphabet $\Sg$ is always assumed to contain the
symbols 0 and 1.
After~\cite{WeihrauchComputAnal00}, 
let us define the \df{wrapping function} $\ig : \Sg^{*} \to \Sg^{*}$ by
 \begin{equation}\label{e.ig}
   \ig(a_{1}a_{2}\dotsm a_{n}) = 110a_{1}0a_{2}0\dotsm a_{n}011.
 \end{equation}
Note that 
 \begin{equation}\label{e.ig-len}
   |\ig(x)| = (2 |x| + 5)\V 6.
 \end{equation}
For strings $x,x_{i} \in \Sg^{*}$, $p, p_{i} \in \Sg^{\og}$, $k \ge 1$,
appropriate tupling functions $\ang{x_{1},\dots,x_{k}}$,
$\ang{x,p}$, $\ang{p,x}$, etc.~can be defined with the help of
$\ang{\cdot,\cdot}$ and $\ig(\cdot)$.

\subsection{Notation and representation}\label{ss.notation-repr}

The concepts of notation and representation, as defined 
in~\cite{WeihrauchComputAnal00}, allow us to transfer
computability properties from some standard spaces to many others.
Given a countable set $C$, a \df{notation} of $C$ is a surjective
partial mapping $\dg :\sbsq \bbN \to C$.
Given some finite alphabet $\Sg \spsq \{0,1\}$ and an arbitrary set $S$,
a \df{representation} of $S$ is a surjective mapping
$\chi :\sbsq \Sg^{\og} \to S$.
A \df{naming system} is a notation or a representation.
Here are some standard naming systems:
 \begin{enumerate}[\upshape 1.]

  \item $\id$, the identity over $\Sg^{*}$ or $\Sg^{\og}$.

  \item $\nu_{\bbN}$, $\nu_{\bbZ}$, $\nu_{\bbQ}$ for the set of natural
numbers, integers and rational numbers.

  \item $\Cf : \Sg^{\og} \to 2^{\bbN}$, the \df{characteristic function
representation} of sets of natural numbers, is defined by
$\Cf(p) = \setof{i : p(i) = 1}$.

  \item $\En : \Sg^{\og} \to 2^{\bbN}$, the \df{enumeration representation} of
sets of natural numbers, is defined by
$\En(p) = \setof{w \in \Sg^{*} : 110^{n+1}11 \ltri p}$.

  \item For $\Dg \sbsq \Sg$, $\En_{\Dg} : \Sg^{\og} \to 2^{\Dg^{*}}$, 
the \df{enumeration representation} of subsets of $\Dg^{*}$, is defined by
$\En_{\Dg}(p) = \setof{w \in \Sg^{*} : \ig(w) \ltri p}$.

 \end{enumerate}

One can define names for all 
computable functions between spaces that are Cartesian 
products of terms of the kind $\Sg^{*}$ and $\Sg^{\og}$.
Then, the notion of computability can be transferred to other spaces as
follows. 
Let $\dg_{i} : Y_{i} \to X_{i}$, $i=1,0$ be naming systems of the spaces
$X_{i}$.
Let $f : \sbsq X_{1} \to X_{0}$, $g : \sbsq Y_{1} \to Y_{0}$.
We say that function $g$ \df{realizes} function $f$ if
 \begin{equation}\label{e.realize}
   f(\dg_{1}(y)) = \dg_{0}(g(y))
 \end{equation}
holds for all $y$ for which the left-hand side is defined.
Realization of multi-argument functions is defined similarly.
We say that a function $f : X_{1} \times X_{2} \to X_{0}$ 
is \df{$(\dg_{1},\dg_{2},\dg_{0})$-computable} if
there is a computable function $g : \sbsq Y_{1} \times Y_{2} \to Y_{0}$
realizing it.
In this case, a name for $f$ is naturally derived from a name of 
$g$.\footnote{Any function $g$ realizing $f$ via~\eqref{e.realize}
automatically has a
certain \df{extensivity} property: if $\dg_{1}(y) = \dg_{1}(y')$ then
$g(y) = g(y')$.}

For representations $\xi,\eta$, 
we write $\xi \le \eta$ if there is a computable function 
$f :\sbsq \Sg^{\og} \to \Sg^{\og}$ with $\xi(x) = \eta(f(x))$.
In words, we say that $\xi$ is \df{reducible} to $\eta$, or that $f$
reduces (translates) $\xi$ to $\eta$.
There is a similar definition of reduction for notations.
We write $\xi \equiv \eta$ if $\xi \le \eta$ and $\eta \le \xi$.

\subsection{Constructive topological space}

\subsubsection{Definitions}
Section~\ref{s.top} gives a review of topological concepts.
A \df{constructive topological space} $\bX = (X, \sg, \nu)$
is a topological space over a set $X$ with a subbase $\sg$ effectively
given as a list $\sg = \{\nu(1),\nu(2),\dots\}$,
and having the $T_{0}$ property (thus, every point is determined uniquely
by the subset of elements of $\sg$ containing it).
By definition, a constructive topological space satisfies the second
countability axiom.\footnote{A constructive topological space
is an effective topological space as defined
in~\cite{WeihrauchComputAnal00}, but, for simplicity
we require the notation $\nu$ to be a total function.}
We obtain a base
  \[
   \sg^{\cap}
 \]
of the space $\bX$ by taking all possible finite
intersections of elements of $\sg$.
It is easy to produce an effective enumeration for $\sg^{\cap}$ from $\nu$.
We will denote this enumeration by $\nu^{\cap}$.

The \df{product operation} is defined over constructive topological spaces
in the natural way.

 \begin{examples}\label{x.constr-topol}\
 \begin{enumerate}[\upshape 1.]

  \item A discrete topological space, where the underlying
set is finite or countably infinite, with a fixed enumeration.
 
  \item\label{i.constr-topol.real}
The real line, choosing the base to be the open intervals
with rational endpoints with their natural enumeration.
Product spaces can be formed
to give the Euclidean plane a constructive topology.

  \item
The real line $\bbR$, with the subbase $\sg_{\bbR}^{>}$ defined as
the set of all open intervals $\opint{-\infty}{b}$ with rational endpoints
$b$.
The subbase $\sg_{\bbR}^{<}$, defined similarly, leads to another topology.
These two topologies differ from each other and from
the usual one on the real line, and they are not Hausdorff spaces.
  
  \item Let $X$ be a set with a constructive discrete topology,
and $X^{\og}$ the set of infinite sequences with elements from $X$,
with the product topology: a natural enumerated basis is also easy to
define.
 
 \end{enumerate}
 \end{examples}

Due to the $T_{0}$ property, every point in our space is determined
uniquely by the set of open sets containing it.
Thus, there is a representation $\gm_{\bX}$ of $\bX$ defined as follows.
We say that $\gm_{\bX}(p) = x$ if
$\En_{\Sg}(p) = \setof{w : x \in \nu(w)}$.
If $\gm_{\bX}(p) = x$ then we say that the infinite sequence
$p$ is a \df{complete name} of $x$:
it encodes all names of all subbase elements containing $x$.
From now on, we will call $\gm_{\bX}$ the \df{complete standard
representation of the space $\bX$}.\footnote{
The book~\cite{WeihrauchComputAnal00} denotes $\gm_{\bX}$ as $\dg'_{\bX}$
instead.
We use $\gm_{\bX}$ only, dispensing with the
notion of a ``computable'' topological space.}

\subsubsection{Constructive open sets, computable functions}
In a constructive topological space $\bX = (X, \sg, \nu)$,
a set $G \sbsq X$ is called \df{r.e.~open} in set $B$ 
if there is a r.e.~set $E$ with 
$G = \bigcup_{w \in E} \nu^{\cap}(w) \cap B$.
It is r.e.~open if it is r.e.~open in $X$.
In the special kind of spaces in which randomness has been developed until
now, constructive open sets have a nice characterization:

 \begin{proposition}\label{p.constr-open-nice-charac}
Assume that the space $\bX = (X, \sg, \nu)$
has the form $Y_{1}\times \dots \times Y_{n}$ where
each $Y_{i}$ is either $\Sg^{*}$ or $\Sg^{\og}$.
Then a set $G$ is r.e.~open iff it is open and the set
$\setof{(w_{1},\dots,w_{n}) : \bigcap_{i}\nu(w_{i}) \sbs G}$ 
is recursively enumerable.
 \end{proposition}
 \begin{proof}
The proof is not difficult, but it relies on the discrete nature of
the space $\Sg^{*}$ and on the fact that the space $\Sg^{\og}$ is compact
and its base consists of sets that are open and closed at the same time.
 \end{proof}

It is easy to see that if two sets are r.e.~open then so is their union.
The above remark implies that a space having
the form $Y_{1}\times \dots \times Y_{n}$ where
each $Y_{i}$ is either $\Sg^{*}$ or $\Sg^{\og}$, also the intersection of
two recursively open sets is recursively open.
We will see that this statement holds, more generally, in all computable
metric spaces.

Let $\bX_{i} = (X_{i}, \sg_{i}, \nu_{i})$ be constructive topological
spaces, and let $f : \sbsq X_{1} \to X_{0}$ be a function.
As we know, $f$ is continuous iff the inverse image $f^{-1}(G)$ of each
open set $G$ is open.
Computability is an effective version of continuity:
it requires that the inverse image
of subbase elements is uniformly constructively open.
More precisely, $f :\sbsq X_{1} \to X_{0}$ is
\df{computable} if the set 
 \[
  \bigcup_{V \in \sg_{0}^{\cap}} f^{-1}(V) \times \{V\}
 \]
is a r.e.~open subset of $X_{1} \times \sg_{0}^{\cap}$.
Here the base $\sg_{0}^{\cap}$ of $\bX_{0}$ is treated as a discrete
constructive topological space, with its natural enumeration.
This definition depends on the enumerations $\nu_{1},\nu_{0}$.
The following theorem (taken from~\cite{WeihrauchComputAnal00})
shows that this computability coincides with the one
obtained by transfer via the representations $\gm_{\bX_{i}}$.
 
 \begin{proposition}\label{p.hertling-computable}
For $i=0,1$, let
$\bX_{i} = (X_{i}, \sg_{i}, \nu_{i})$ be constructive topological spaces.
Then a function $f :\sbsq X_{1} \to X_{0}$ is
computable iff it is $(\gm_{\bX_{1}},\gm_{\bX_{0}})$-computable for the 
representations $\gm_{\bX_{i}}$ defined above.
 \end{proposition}

As a name of a computable function, we can use the name of the enumeration
algorithm derived from the definition of computability, or the name
derivable using this representation theorem.

 \begin{remark}
As in Proposition~\ref{p.constr-open-nice-charac},
it would be nice to have the following statement, at least for total
functions: 
``Function $f : X_{1} \to X_{0}$ is computable iff the set
 \[
 \setof{(v, w) : \nu^{\cap}_{1}(w) \sbs f^{-1}[\nu_{0}(v)] }
 \]
is recursively enumerable.''
But such a characterization seems to require compactness and possibly more.
  \end{remark}

Let us call two spaces $X_{1}$ and $X_{0}$ \df{effectively homeomorphic}
if there are computable maps between them that are inverses of each
other.
In the special case when $X_{0}=X_{1}$, we say 
that the enumerations of subbases
$\nu_{0},\nu_{1}$ are \df{equivalent} if the identity
mapping is a effective homeomorphism.
This means that there are recursively enumerable sets $F,G$ such that
 \[
  \nu_{1}(v) = \bigcup_{(v, w) \in F} \nu_{0}^{\cap}(w) \txt{ for all $v$},
\quad
  \nu_{0}(w) = \bigcup_{(w, v) \in G} \nu_{1}^{\cap}(v) \txt{ for all $w$}.
 \]
Lower semicomputability is a constructive version of lower
semicontinuity.
Let $\bX = (X, \sg, \nu)$ be a constructive topological space.
A function $f :\sbsq X \to \ol\bbR_{+}$ is called \df{lower semicomputable}
if the set $\setof{(x,r): f(x) > r}$ is r.e.~open.
Let $\bY = (\ol\bbR_{+}, \sg_{\bbR}^{<}, \nu_{\bbR}^{<})$ be the effective
topological space introduced in
Example~\ref{x.constr-topol}.\ref{i.constr-topol.real},
in which $\nu_{\bbR}^{>}$ is an enumeration of all open intervals of the
form $\rint{r}{\infty}$ with rational $r$.
It can be seen that $f$ is lower semicomputable iff it is
$(\nu,\nu_{\bbR}^{>})$-computable.

\subsubsection{Computable elements and 
sequences}\label{sss.computable-elements}
Let $\bU = (\{0\}, \sg_{0}, \nu_{0})$ 
be the one-element space turned into a trivial constructive
topological space, and let $\bX = (X, \sg, \nu)$ be another constructive
topological space.
We say that an element $x \in X$ is \df{computable} if the function
$0 \mapsto x$ is computable.
It is easy to see that this is equivalent to the requirement that
the set $\setof{u : x \in \nu(u)}$ is recursively enumerable.
Let $\bX_{j}= (X_{j}, \sg_{j}, \nu_{j})$, 
for $i=0,1$ be constructive topological spaces.
A sequence $f_{i}$, $i=1,2,\dots$ of functions with
$f_{i} : X_{1} \to X_{0}$ is a
\df{computable sequence of computable functions} if 
$(i, x) \mapsto f_{i}(x)$ is a computable function.
Using the s-m-n theorem of recursion theory, it is easy to see that this
statement is equivalent to the statement that there is a recursive function
computing from each $i$ a name for the computable function $f_{i}$.
The proof of the following statement is not difficult.

 \begin{proposition}\label{p.one-arg-cpt}
Let $\bX_{i} = (X_{i}, \sg_{i}, \nu_{i})$ 
for $i=1,2,0$ be constructive topological spaces, and
let $f: X_{1} \times X_{2} \to X_{0}$, and assume that $x_{1} \in X_{1}$ is
a computable element.
 \begin{enumerate}[\upshape 1.]

  \item If $f$ is computable and 
then $x_{2} \mapsto f(x_{1}, x_{2})$ is also computable.

  \item If $\bX_{0} = \ol\bbR$, and $f$ is lower semicomputable
then $x_{2} \mapsto f(x_{1}, x_{2})$ is also lower semicomputable.

 \end{enumerate}
 \end{proposition}

\subsection{Computable metric space}

Following~\cite{BrattkaPresserMetric03}, we define
a computable metric space as a tuple $\bX = (X, d, D, \ag)$ where $(X,d)$
is a  metric space, with a countable dense subset $D$
and an enumeration $\ag$ of $D$.
It is assumed that the real function $d(\ag(v),\ag(w))$ is computable.
As $x$ runs through elements of $D$ and $r$ through positive rational
numbers, we obtain the enumeration of
a countable basis $\setof{B(x, r) : x \in D, r\in \bbQ}$ (of balls or radius $r$
and center $x$) of $\bX$,  
giving rise to a constructive topological space $\tilde\bX$.
Let us call a sequence $x_{1}, x_{2},\dots$ a \df{Cauchy} sequence if
for all $i<j$ we have $d(x_{i},x_{j}) \le 2^{-i}$.
To connect to the type-2 theory of computability developed above,
the \df{Cauchy-representation} $\dg_{\bX}$ of the space can be defined in a
natural way.
It can be shown that as a representation of $\tilde\bX$, it is equivalent
to $\gm_{\tilde\bX}$: $\dg_{\bX} \equiv \gm_{\tilde\bX}$.

\begin{example}\label{x.cptable-metric-{0}{1}}
Example~\protect\ref{x.Cclint{0}{1}} is a computable
metric space, with either of the two (equivalent)
choices for an enumerated dense set.
\end{example}

Similarly to the definition of a computable sequence of computable
functions in~\ref{sss.computable-elements}, we can
define the notion of a computable sequence of bounded computable functions,
or the computable sequence $f_{i}$ of computable Lip\-schitz functions: 
the bound and the Lip\-schitz constant of $f_{i}$ are required to be
computable from $i$.
The following statement shows, in an effective form,
that a function is lower semicomputable if and only if it is the supremum
of a computable sequence of computable functions.

 \begin{proposition}\label{p.lower-semi-as-limit}
Let $\bX$ be a computable metric space.
There is a computable mapping that to each name of a nonnegative
lower semicomputable
function $f$ assigns a name of a computable sequence of computable 
bounded Lip\-schitz functions $f_{i}$ whose supremum is $f$.
 \end{proposition}
\begin{proof}[Proof sketch]
Show that $f$ is the supremum of a computable sequence of computable
functions $c_{i} 1_{B(u_{i}, r_{i})}$ where $u_{i}\in D$ and 
$c_{i}, r_{i} > 0$ are rational.
Clearly, each indicator function $1_{B(u_{i},r_{i})}$ is the supremum
of a computable sequence of computable functions $g_{i,j}$.
We have $f = \sup_{n} f_{n}$ where $f_{n} = \max_{i \le n} c_{i} g_{i,n}$.
It is easy to see that the bounds on the functions $f_{n}$ are computable
from $n$ and that they all are in $\Lip_{\bg_{n}}$ for a
$\bg_{n}$ that is computable from $n$.
 \end{proof}

The following is also worth noting.

 \begin{proposition}
In a computable metric space, the intersection of two r.e.~open sets is
r.e.~open.
 \end{proposition}
 \begin{proof}
Let $\bg = \setof{B(x, r) : x \in D, r\in \bbQ}$ be a basis of our space.
For a pair $(x,r)$ with $x \in D$, $r \in \bbQ$, let
 \[
  \Gg(x,r) = \setof{(y,s): y\in D,\;s\in \bbQ,\; d(x,y)+s < r}.
 \]
If $U$ is a r.e.~open set, then there is a r.e.~set
$S_{U} \sbs D \times \bbQ$ with $U = \bigcup_{(x,r) \in S_{U}} B(x,r)$.
Let $S'_{U} = \bigcup\setof{\Gg(x,r) : (x,r) \in S_{U}}$, then we have
$U = \bigcup_{(x,r) \in S'_{U}} B(x,r)$.
Now, it is easy to see 
 \[
  U\cap V = \bigcup_{(x,r) \in S'_{U} \cap S'_{V}} B(x,r).
 \]
 \end{proof}


\providecommand{\bysame}{\leavevmode\hbox to3em{\hrulefill}\thinspace}
\providecommand{\MR}{\relax\ifhmode\unskip\space\fi MR }
\providecommand{\MRhref}[2]{%
  \href{http://www.ams.org/mathscinet-getitem?mr=#1}{#2}
}
\providecommand{\href}[2]{#2}

\end{document}